\documentclass[preprint2]{aastex631}
\usepackage[utf8]{inputenc}
\usepackage{t1enc}
\usepackage{amssymb}
\usepackage{graphicx}
\usepackage{amsmath}
\usepackage{natbib}
\usepackage{booktabs}
\usepackage{rotating}
\usepackage{hyperref}
\usepackage[hyphens]{url}
\usepackage{silence}
\usepackage{etoolbox}
\usepackage{supertabular}
\usepackage{soul}
\usepackage{txfonts}

\begin{document}
\sloppy

\title{Solar System objects observed with TESS -- Early Data Release 2: \\
I. Spin-shape recovery potential of multi-epoch TESS observations}
\shorttitle{TSSYS-EDR2: Multi-epoch observations of Solar System objects by TESS}

\author[0009-0008-2021-1098]{N\'ora Tak\'acs}
\affiliation{Konkoly Observatory, HUN-REN Research Centre for Astronomy and Earth Sciences, Konkoly Thege 15-17, H-1121~Budapest, Hungary}
\affiliation{CSFK, MTA Centre of Excellence, Budapest, Konkoly Thege 15-17, H-1121, Hungary}
\affiliation{ELTE E\"otv\"os Lor\'and University, Institute of Physics and Astronomy, Budapest, Hungary}

\author[0000-0002-8722-6875]{Csaba Kiss}
\affiliation{Konkoly Observatory, HUN-REN Research Centre for Astronomy and Earth Sciences, Konkoly Thege 15-17, H-1121~Budapest, Hungary}
\affiliation{CSFK, MTA Centre of Excellence, Budapest, Konkoly Thege 15-17, H-1121, Hungary}
\affiliation{ELTE E\"otv\"os Lor\'and University, Institute of Physics and Astronomy, Budapest, Hungary}

\author[0000-0002-1698-605X]{R\'obert Szak\'ats}
\affiliation{Konkoly Observatory, HUN-REN Research Centre for Astronomy and Earth Sciences, Konkoly Thege 15-17, H-1121~Budapest, Hungary}
\affiliation{CSFK, MTA Centre of Excellence, Budapest, Konkoly Thege 15-17, H-1121, Hungary}

\author[0000-0001-5449-2467]{Andr\'as P\'al}
\affiliation{Konkoly Observatory, HUN-REN Research Centre for Astronomy and Earth Sciences, Konkoly Thege 15-17, H-1121~Budapest, Hungary}
\affiliation{CSFK, MTA Centre of Excellence, Budapest, Konkoly Thege 15-17, H-1121, Hungary}
\affiliation{ELTE E\"otv\"os Lor\'and University, Institute of Physics and Astronomy, Budapest, Hungary}

\begin{abstract}

Using multidirectional measurements from the Transiting Exoplanet Survey Satellite (TESS), we investigated the viability of determining the approximate shape and spin axis orientations for 44 selected main belt asteroids, using light curve inversion, assuming Lommel-Seeliger ellipsoids. This study aims to investigate the applicability of low-degree-of-freedom shape models in those cases when rotation periods can be accurately determined, but light curves are only available in a limited number of geometries or orbital phases. Our results are compared with the shape and spin axis solutions obtained for the same set of asteroids by more complex light curve inversion methods using mainly ground-based measurements, available via the Database of Asteroid Models from Inversion Techniques (DAMIT).The best-fit spin-axis orientations show a moderately good match with the DAMIT solutions; however, a better agreement is reached with triaxial ellipsoid solution obtained from other large, independent surveys. This suggests that while TESS-only data works well for finding rotation periods, it has its limitations when determining asteroid shape and spin-axis orientation. We discuss the challenges and potential applications of this approach for studying large number of asteroids observed by TESS.
\end{abstract}


\section{Introduction}
\label{sec:introduction}
Light curve inversion techniques saw a remarkable development in the last decades \citep{2015aste.book..183D}. The main outcomes of these models are the shape and spin axis orientation of asteroids. On one hand, they are important for individual targets as e.g. the Yarkovsky and YORP effects \citep{2015aste.book..509V} can significantly modify the orbit and the spin state of these asteroids. Shapes and spin axis orientations, when looked at the level of populations, on the other hand, hold information on the formation and collisional evolution of the minor bodies in our Solar System \citep[see e.g.][]{Hanus2023}. In most cases, light curve inversion techniques require the collection of asteroid light data through several years or even decades to cover the necessary range of orbital phases for a meaningful shape/spin axis recovery. This usually leads to a very inhomogeneous dataset obtained with various -- usually ground-based -- telescopes that often consist of single brightness measurements or short photometry series. 

In the last decade, large surveys allowed the determination of spin states and shapes for a large number of asteroids, as it was obtained e.g. by \citet{Durech2023} for 8600 asteroids using Gaia DR3 data. While shape models can be quite complex, the triaxial ellipsoid model has been found to give satisfactory results in the case of a limited set of measurements, providing pole orientation, rotation period, and ellipsoid axis ratios. This is also true for large surveys with sparse data, such as Gaia, as demonstrated by \citet{SantanaRos2015}. In a recent work \citet[][referred to as C24 throughout this paper]{Cellino2024} determined spin and shape properties for 8678 asteroids from Gaia DR3 photometry, assuming simple triaxial ellipsoid shape models.
In these models, surface scattering can also be taken into account, e.g. using the Lommel-Seeliger scattering ellipsoidal asteroids \citep{ML2015,Muinonen2015}.

The Transiting Exoplanet Survey Satellite \citep[TESS,][]{2015JATIS...1a4003R, 2015JATIS...1a4003R} has proven to be a highly efficient tool for observing small Solar System bodies, offering a unique advantage in the field of asteroid shape modelling. Unlike previous missions such as Kepler/K2 \citep{10.1088/2514-3433/ab9823ch5, howell2014}, which primarily provided data for pre-selected targets \citep{pal2015,pal2016,farkas2017,kiss2017} or through limited superstamps \citep[see e.g.][]{kiss2016,szabo2016,molnar2018}, TESS provides extensive, uninterrupted photometric time series (up to 28 days) data across multiple, typically 3-4 sectors. This enables long-term monitoring, a diverse sampling of observational geometries, and accurate determination of rotational periods for tens or even hundreds of thousands of asteroids, comets, and trans-Neptunian objects. These observations, though serendipitous, yield light curves of exceptional quality that can potentially be used to explore the shape and spin axis orientation of these objects. \citep{pal2018,holman2019,payne2019,mcneill2019,pal2020,rice2020,woods2021}.

TESS is nearing the end of its sixth year of monitoring, providing multiple observations of the same targets over a long period of time. This allows us to view objects from different spatial orientations, theoretically enabling the recovery of their shapes and spin-axis orientations simultaneously -- see also the TESS Observations web page\footnote{https://tess.mit.edu/observations/} for more details about the instrument. Due to its high cadence, TESS can eliminate the uncertainty in the rotation periods, allowing us to confidently confirm periods in the $\sim$2\,h -- 2 weeks range, potentially including many slow rotators. The underlying photometric database is also highly homogeneous, with far fewer biases than in ground-based observations \citep{pal2018,pal2020}.

In this paper, we investigate the applicability of a simple method to obtain triaxial ellipsoid shape models and spin axis orientations for asteroids observed with TESS. 
We assess the suitability of TESS's typical sampling for determining these basic parameters, identify the method's limitations, and determine whether meaningful statistical insights can be derived from TESS data alone. We analyse 44 asteroids, focusing on validating the use of pure TESS data to recover their fundamental shape and pole-axis characteristics. This validation is carried out by comparing our results with existing ground-based, as well as space projects and databases \citep{durech2010,2020A&A...643A..59D}, in particular the solutions listed in the Database of Asteroid Models from Inversion Techniques (DAMIT)\footnote{https://astro.troja.mff.cuni.cz/projects/damit/}. 

The paper is structured as follows: In Section \ref{sec:objectindexing}, we explain the hierarchy of databases designed for fast lookup and selection of asteroids observed by TESS, using various criteria. Section  \ref{sec:photometry} provides a brief overview of the photometry process for our targets of interest, highlighting algorithmic and implementation improvements compared to the first data release \citep{pal2020}. Section  \ref{sec:targets} outlines how we integrated our databases with existing ones \citep{durech2010, 2020A&A...643A..59D} to identify validation targets. Our results are presented in Section \ref{sec:results}, followed by the discussion in Section \ref{sec:summary}.

\section{Object indexing and back-end catalogues}
\label{sec:objectindexing}

In order to provide an efficient system for querying asteroid observations throughout the TESS field of views, we developed and employed two back-end software stacks. The first one, \texttt{EPHEMD}, is a server-client architecture system, providing both querying and ephemerides computation based on the retrieval of the official ephemerides of \texttt{MPCORB.dat.gz} as provided by the Minor Planet Center\footnote{https://minorplanetcenter.net/data}. The conceptual design of the \texttt{EPHEMD} solution is described in \cite{pal2018} and \cite{pal2020}, while it was also employed earlier in another space-borne asteroid photometry projects, both in the optical \citep{szabo2016,molnar2018} and far-infrared regime \citep{szakats2017,szakats2020}.

While having the \texttt{EPHEMD} stack as the primary back-end for executing generic visibility queries and ephemerides computation, we built a look-up database of TESS-specific observation constraints. This database is implemented in SQL and formed around the [sector, camera, CCD, target] unique tuples. For each tuple, the total number of observations (i.e., when the target was within the field of view and was detected by the silicon-based CCD sensors), the cadence and the total number of frames available are associated while the means of the expected brightness, heliocentric and observer-centric distances, and the apparent speed in the units of pixel/cadence are computed and stored. In Table~\ref{tab:obslist}, the [sector, camera, CCD, target] tuples are listed using the visibility query functionality of \texttt{EPHEMD} aided with the astrometric plate solutions for the TESS frames \citep[see e.g. Sec.~3.2 in][]{pal2020} while the means of the aforementioned quantities are obtained via the ephemerides functionality of \texttt{EPHEMD}.

Luckily, the optimal indexing of the SQL tables is rather straightforward: once an SQL index is assigned to the object identifier (i.e. minor planet number, provisional designation or comet name), the sector number and the average expected brightness (magnitude) of the object, most of the required queries (needed by the current study) are executed rather fast and efficient. We note here that there is no need for a unique index (or primary key) on the 4-tuples formed by the target and the sector/camera/CCD fields. In addition, the access of the database remains smooth while expanding it with object visibility information from any further TESS sectors. Querying this database can be done either directly from an SQL client or through small wrapper scripts, which provide additional functionality and filtering that would be less efficient if fully implemented within the SQL schema.

\begin{table*}[!ht]
\caption{Summary of the observations used in this paper, shown with the first three asteroids as an example. The S/C/C fields denote the TESS sector, camera and CCD identifier, Start and End indicate the first and last Julian Dates of the given sector's observation, respectively. $q$ is the fraction of the observed asteroid track with respect to the total observing time for the particular sector (i.e. $q=1$ is equivalent to approximately $25$\, days). $m_{\rm V}$ is the predicted visual brightness of the object while $(\lambda,\beta)$ are the apparent longitude and latitude in the J2000 ecliptic reference frame at the centre of the observed arc. The whole table is available in electronically readable format.}
\label{tab:obslist}

\begin{center}
\begin{tabular}[t]{rrcccrrr}
Asteroid & S/C/C & Start & End & $q$ & $m_{\rm V}$ & $\lambda$ & $\beta$  \\
 & & [JD] & [JD] & & [mag] & [deg] & [deg] \\
\hline
  22  & 29/1/2 & 2459088.24678 & 2459107.19816 & 0.73 & 10.7 & 348.9 &-20.6 \\
      & 43/4/4 & 2459474.17023 & 2459491.92022 & 0.72 & 11.7 & 98.0  &  2.5  \\
      & 44/3/4 & 2459500.79521 & 2459524.43409 & 0.98 & 11.3 & 103.6 &  4.9 \\
\hline
  94  &  1/1/3 & 2458325.32352 & 2458337.51101 & 0.45 & 12.3 & 317.2 & -8.6 \\
      & 18/1/3 & 2458790.67749 & 2458793.53165 & 0.33 & 11.6 &  41.1 &  7.8 \\
      & 46/4/3 & 2459552.84381 & 2459560.21186 & 0.32 & 14.0 & 202.5 & -0.7\\
\hline
  156 & 17/1/4 & 2458772.19833 & 2458777.73999 & 0.74 & 13.5 &  29.0 &  8.0  \\
      & 33/1/4 & 2459201.73979 & 2459213.26061 & 0.83 & 13.0 &  90.2 & -7.4  \\
      & 46/4/3 & 2459552.95492 & 2459570.10074 & 0.69 & 13.6 & 199.3 & -7.9 \\

\end{tabular}
\end{center}
\end{table*}

\section{Photometric algorithms and post-processing}
\label{sec:photometry}
 
Most of the steps related to the photometric processing have been performed the same way as it is described in \cite{pal2020}. Since the total amount of data volume available for processing increased more than ten-fold compared to the first data release of asteroid photometry (due to the reduction of the observing cadence of full-frame images from 30 minutes down to 10 minutes starting from the third year of TESS operations and, of course, the inclusion of almost three years of additional data) and having the five consecutive Ecliptic sectors of 42, 43, 44, 45 and 46, several optimizations have also been implemented at the lowest levels of the FITSH package tasks \citep{pal2012} while keeping the same functionality. In addition, the higher-level processing steps have also been tuned in many aspects as we summarize in the following list below.

\begin{itemize}
\item The selection criteria for the {\it individual median reference frames} \citep[IMRFs, see][]{pal2020} have been improved by considering visual verification of the frames to avoid the images affected by the so-called ``firefly'' phenomena \citep{villasenor2019}. 
\item Proper masking has been implemented to avoid the vicinity of the brightest stars having a prominent undershoot around the saturated pixels \citep[see e.g. Fig.~6.11 in][]{tenenbaum2018} and ghosts caused by intra-camera cross-talks \citep[Fig.~6.13 in][]{tenenbaum2018}. These effects would imply a non-linear response during convolution, image subtraction and unstrapping.
\item The overall aperture sizes are reduced in order to increase the signal-to-noise ratio for the fainter targets while keeping larger apertures for the brighter ones.
\item Automatic masking procedures are included to exclude points when an asteroid crosses apparently the path of another asteroid (which is rather frequent in the Ecliptic  Sectors of $42-46$) as well as in the case of known variable stars. 
\end{itemize}

This efficient processing allowed the creation of light curves with multiple apertures, enabling the selection of the aperture providing the highest overall signal-to-noise ratio for any of the targets. The total volume of pure asteroid light curve data is around $\sim$490\,GiB ($\sim$527\,GB) up to Sector 48 (the last sector considered in this paper), which is equivalent to $\sim$3.23 billion of individual photometric data points for the $\sim$1.76 million unique [sector, camera, CCD, target] tuples, from where the number of unique [target] fields is around $\sim$750 thousand.  Of course, these database entries include the sub-detection photometry for fainter targets below the individual per-frame detection limit. However, keeping such light curves (which are indeed dominating the database) is also helpful since the linearity of the photometric flux extraction process allows us to average out the individual sub-detection photometric fluxes for a significantly positive average flux on longer timescales. 

The full process, along with access to the database will be provided by a separate paper. In the following sections of this paper, we focus only on the targets that are selected for validating the spin-axis determination algorithms purely from TESS data. This is important because all of the results provided later in this paper do not exploit any properties of these objects, i.e. the full data processing pipeline does not treat them as special in any sense.

\section{Target selection and shape modelling}
\label{sec:targets}

\subsection{Target selection and period determination}
\label{sec:selection}

The target asteroids of this paper have been selected using the following criteria: 
1) the target had to be measured by TESS in at least 3 relatively distant (not adjacent) sectors, necessary for meaningful shape recovery; 2) its apparent magnitude should be at least 18.0; and 3) it should already have a shape and spin orientation solution in the DAMIT catalogue.

Using these criteria and the asteroid search query mentioned in Sect. \ref{sec:objectindexing}., we obtained a total of 44 targets with a predicted visual magnitude between 10.7 and 17.9. The list of selected asteroids and their observational characteristics can be found in Table~\ref{tab:obslist}. 

In the light curves of these 44 asteroids, outliers were identified and filtered before further processing. These outliers may have occurred due to events such as the target asteroid passing in front of a variable star, or a close encounter with another asteroid. A data point was considered an outlier if its deviation from the mean, divided by the standard deviation, was greater than 3 sigma.
After removing outliers, we performed a period search on the light curves. This was achieved through frequency scanning, which produced residual spectra. In each case, the synodic period and its uncertainty were determined by identifying the highest signal-to-noise peak and fitting a Gaussian function to it (for details on the frequency spectra generation process, see \citealp{pal2020}).

\begin{figure*}[ht!]
    \centering
    \includegraphics[width=\textwidth]{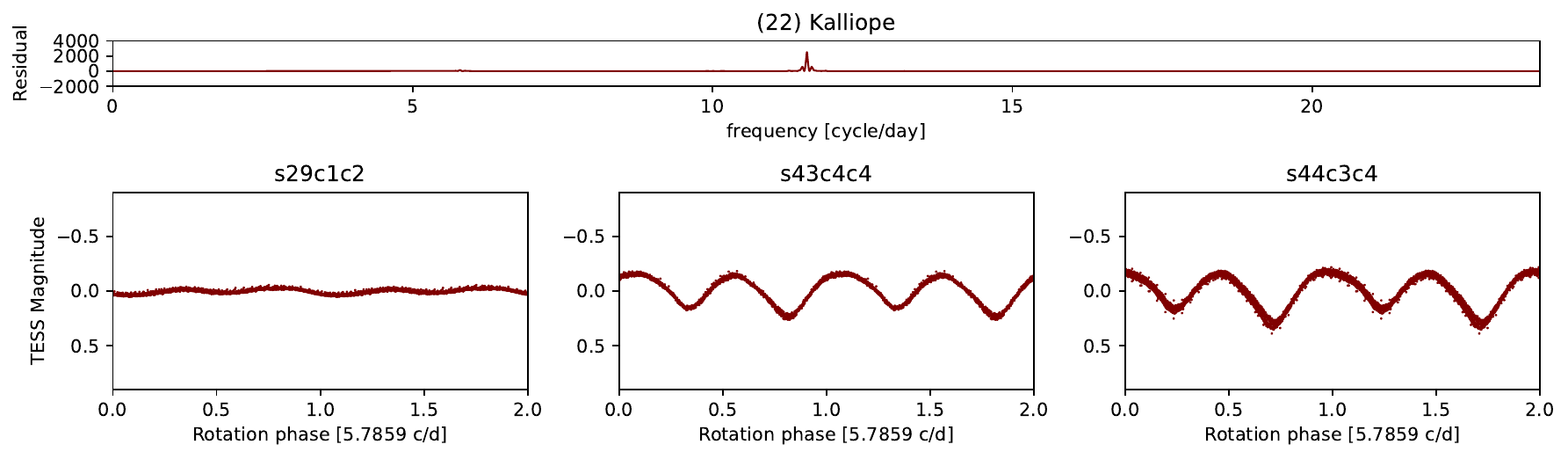}
    \includegraphics[width=\textwidth]{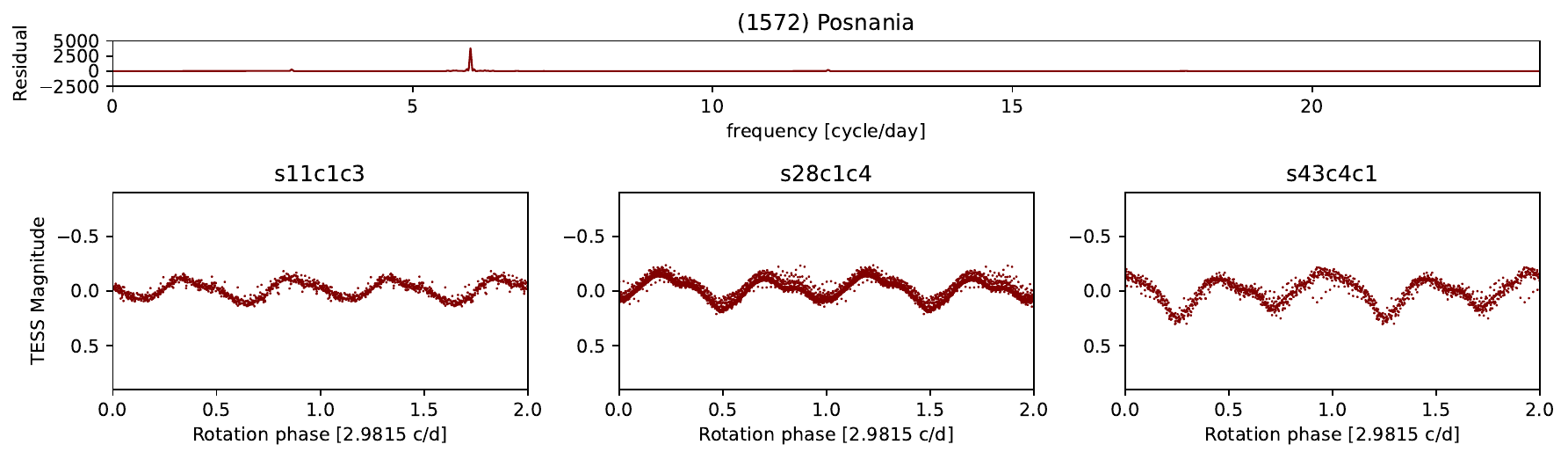}
    \includegraphics[width=\textwidth]{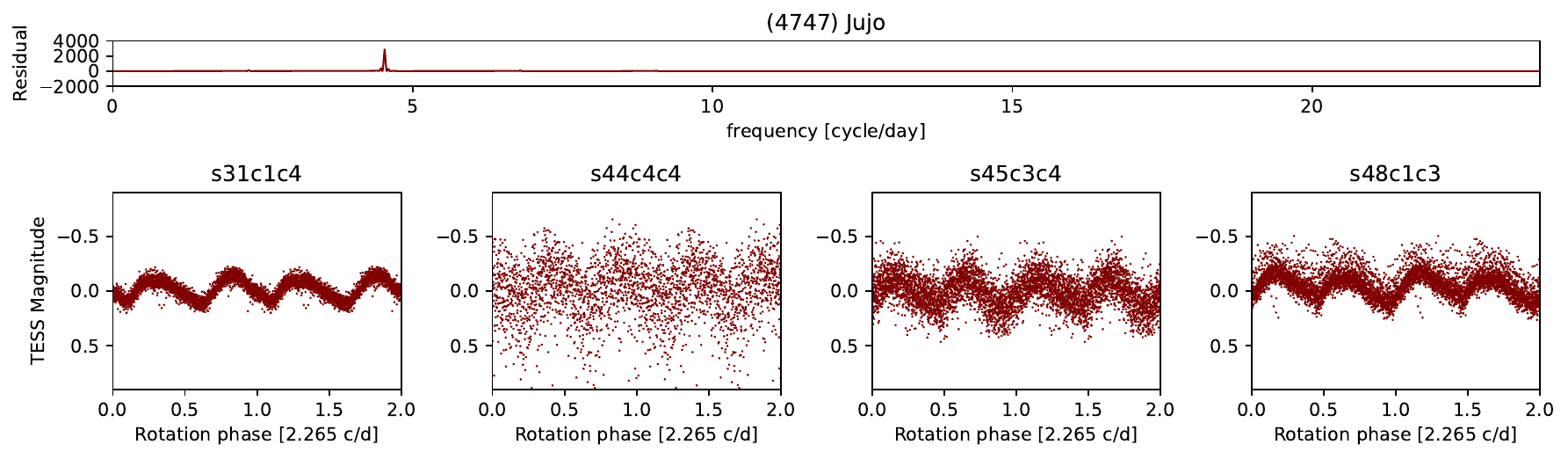}
    \caption{Frequency spectrum and folded light curves for three asteroids. The top panel (first and second rows) shows the frequency spectrum of (22) Kalliope from TESS sector 43, and its folded light curves from sectors 29, 43 and 44, with an average brightness of 10.5 magnitudes over the sectors. The middle panel (third and fourth rows) shows the same for (1572) Posnania, based on sector 28 for the frequency spectrum and sectors 11, 28 and 43 for the folded light curves, where the average brightness is 14.1 magnitudes. The bottom panel (fifth and sixth rows) shows the results for (4747) Jujo, with the frequency spectrum from sector 31 and folded light curves from sectors 31, 44, 45 and 48, where the average magnitude is 16.7.}
    \label{fig:foldedlcs22}
\end{figure*}

Examples of folded light curves, obtained by TESS, and their frequency spectra are shown in Fig.~\ref{fig:foldedlcs22}. The three asteroids presented as examples span different magnitude regimes, where the Flux from the TESS magnitude can be estimated using the following equation (for further details, see the TESS Instrument Handbook)\footnote{https://archive.stsci.edu/files/live/sites/mast/files/home/
missions-and-data/active-missions/tess/\_documents/TESS
\_Instrument\_Handbook\_v0.1.pdf}

\begin{equation}
    Flux \, [erg/s] = 10^{((20.44 - T_{mag})/2.5)} \, .
\end{equation}

This procedure was applied to all 44 asteroids across all available measurements and aperture settings for each observation. The best aperture was selected by comparing the signal-to-noise ratios of the light curves, folded with the same, dominant frequency. TESS light curve data for these 44 asteroids are also available in electronic format.\footnote{https://cloud.konkoly.hu/s/EZ6eryAn6M2Msap}

\subsection{Shape and spin pole modelling}
\label{sec:modelling}

The simplest asteroid shape model which shows brightness variations, with a homogeneous albedo distribution on the surface, has the shape of a triaxial ellipsoid. Here, we refer to this model as the ``simplest'' because all of the shape models expressed as the lowest possible (i.e. second-order) spherical harmonics expansion are equivalent to a tri-axial ellipsoid with some orientation parameter of its pre-defined axis - which can also be the rotation axis of the body. 
In our triaxial ellipsoid model, the three semi-axes can be simply parametrised as 
$a=1+x$, $b=1$ and $c=1-x$, where we assumed the $x$ parameter ranging from 0.0 to 0.5.

In this model, the variations in the observed light curve can be explained by the changing apparent cross-section, a concept known as the amplitude method, as described by \citet{1986Icar...68....1M}. Additionally, the solar illumination and observer vectors relative to the surface normal are considered, integrating these effects over the illuminated fraction of the body as seen from the observer’s perspective, taking into account Lommel-Seeliger scattering \citep{Muinonen2015,ML2015}. These variations depend on the axial ratios (b/a and c/a, hence $x$ in our case) of the ellipsoid and the orientation of the asteroid’s spin axis ($\lambda_p$, $\beta_p$ ecliptic coordinates), and the best-fit model is obtained by $\chi^2$ minimization in the \{$x$, $\lambda_p$, $\beta_p$\} space. 

\section{Results}
\label{sec:results}

For all 44 asteroids, $\chi^2$ maps represented in ecliptic coordinates using HEALPix\footnote{http://healpix.sourceforge.net} \citep{Zonca2019, 2005ApJ...622..759G} projections summarize our pole solution results. We use a HEALPix resolution of \(\text{NSIDE} = 20\) (N$_{\rm{pix}}$ = 4800 ) and a Mollweide projection for visualization (see Figs.~4--6 in the Appendix as examples; all other figures are presented in electronic format).

For each object, we generated two sets of $\chi^2$ maps to visualize the distributions of the pole solution.
In the first set, white pixels indicate the regions where the $\chi^2$ values are within the lowest 2\% of the maximum value, highlighting the most probable pole solutions.
In the second set, we applied a different selection criterion by identifying a fixed number of the lowest $\chi^2$ values. Specifically, we selected 96 pixels from the total of 4800 pixels in the HealPix grid used for mapping the pole orientations. This means that, similar to the first set, approximately 2\% of the entire mapped sky was identified as the most probable region for each object.

The red dot in each figure indicates the 'center of mass' of the clustered white areas. The pole orientation results provided by DAMIT are marked with white circles, enabling a direct comparison between the results derived from the simple shape model based on TESS data and those obtained from a more sophisticated modelling approach using ground-based measurements. The $\chi^2$ values are normalised between 0 and 1 in each figure for easier visualisation. In Table~\ref{tab:Results}, we summarise our results for each asteroid, based on the lowest 2\% of the maximum $\chi^2$ values, along with a comparison to results from the DAMIT database and \citet{Cellino2024} who determined spin pole and shape solutions for 8678 asteroids assuming triaxial ellipsoid shapes using Gaia data.
Since both our method and the DAMIT database give several results for the pole orientation of an object, we always take the angular distance of the closest of these results. As our shape solutions are not directly comparable with the complex DAMIT shape solutions, we concentrate on comparing the pole solutions provided by the different methods. Following the example of \cite{Hanus2023} and \cite{2018A&A...617A..57D}, differences of more than 30 degrees between solutions should be considered as a lack of agreement.

Table~\ref{tab:Results}. also presents the rotational periods and the shape parameter $x$. The synodic periods generally show a good agreement with the literature values, except for (13289) 1998 QK75, where the discrepancy was significant. The period was determined by \citep{2019A&A...631A...2D} to be $43.2618$ hours, while our analysis resulted in a period of $463.89$ hours. The light curves of this object, folded with the period we have determined are also included in the online available database. 
Regarding the shape parameter, in most cases, reaches the upper limit of $0.5$. This corresponds to an unusually elongated shape ($a/c=3$), which is rarely observed among asteroids. This result is likely due to the limited range of viewing geometries in the TESS dataset, which may not provide enough constraints for a reliable shape determination.

Within our sample, two objects - (1381) Danubia and (4717) Kaneko - have significantly larger uncertainties in their beta coordinates, with values of $56.44^{\circ}$ and $40.78^{\circ}$ respectively. This is probably due to the specific viewing geometries provided by TESS, where the asteroid was observed from directions that resulted in relatively small amplitude variations between sectors. In contrast, the DAMIT solutions were able to obtain more accurate spin axis determinations due to the more widespread orbital coverage.

\begin{table*}[!ht]
\scriptsize
\begin{center}
\begin{tabular}{r|rrrrrrrrrrrrrr} 
\toprule 
\textbf{ID} & $\mathbf{P}$  & $\mathbf{\lambda_{T}}$ & $\mathbf{\delta\lambda_{T}}$ & $\mathbf{\beta_{T}}$  & $\mathbf{\delta\beta_{T}}$ & $\mathbf{A}$ & $\mathbf{\lambda_{D}}$  & $\mathbf{\beta_{D}}$ & Ref. & \textbf{d} &\textbf{p}  & $\mathbf{\Xi}$  & \textbf{d}$\mathbf{_{C24}}$& \textbf{X} \\ 
      \hspace{5mm} & \textbf{[h]}  & \textbf{[deg]} & \textbf{[deg]} &\textbf{[deg]}  &\textbf{[deg]} &\textbf{[deg$^\mathbf{2}$]} &\textbf{[deg]}  &\textbf{[deg]} & & \textbf{[deg]} &  & \textbf{[}$\mathbf{\%}$\textbf{]} & \textbf{[deg]}&  \\ 
\hline 
\hline 
\textbf{22} & 4.15 & 188 & 2.45 & -13 & 9.57 & 12.90 & 196 & 2 & H17 & 16.61  &0.042 & 0.26 & & 0.45 \\ 
 \hline 
\textbf{94} & 7.22 & 354 & 1.13 & 18 & 1.01 & 0.10& 55 & 11 & H17 & 58.86 & 0.483 & 0.94 & & 0.05 \\ 
 \hline 
\textbf{156} & 22.15 & 208 & 3.99 & 35 & 1.60 & 2.00 & 197 & 9 & Ď20 & 27.48 & 0.113 & 0.64 & & 0.30 \\ 
 \hline 
\textbf{292} & 8.93 & 57 & 11.34 & 77 & 2.63 & 4.00 & 110 & 39 & Ď20 & 44.12 &  0.282 & 0.55 && 0.50 \\ 
 \hline 
\textbf{328} & 10.98 & 39 & 2.65 & -26 & 3.56 & 5.10 & 31 & 11 & Ď19 & 37.57 & 0.207 & 0.27 & & 0.35 \\ 
 \hline 
\textbf{424} & 40.11 & 176 & 5.11 & -59 & 5.28 & 5.00 & 172 & -6 & Ď20 & 53.31 & 0.402 & 0.86 & & 0.25 \\ 
 \hline 
\textbf{645} & 53.49 & 74 & 22.01 & 59 & 9.55 & 21.80 & 162 & 58 & Ď20 & 42.46 & 0.262 & 0.21 & & 0.25 \\ 
 \hline 
\textbf{844} & 6.78 & 256 & 43.44 & 74 & 6.56 & 16.05 & 302 & 68 & Ď16 & 15.51 & 0.036 & 0.15 & & 0.20 \\ 
 \hline 
\textbf{950} & 211.70 & 94 & 14.48 & 59 & 7.65 & 30.95 & 10 & 89 & Ď19 & 30.69 &0.140 & 0.52 & & 0.40 \\ 
 \hline 
\textbf{1182} & 29.86 & 58 & 20.40 & -76 & 5.40 & 11.50 & 34 & -42 & Ď18 & 35.48 & 0.186 & 0.59 & & 0.50 \\ 
 \hline 
\textbf{1381} & 5.26 & 196 & 19.62 & 8 & 56.44 & 67.50 & 174 & 30 & Ď20 & 30.25 & 0.136 & 0.49 &38.76& 0.50 \\ 
 \hline 
\textbf{1508} & 9.19 & 257 & 0.61 & 65 & 1.20 & 1.00 & 166 & 73 & H16 & 30.00 & 0.134 & 0.72 & & 0.50 \\ 
 \hline 
\textbf{1572} & 8.05 & 155 & 11.57 & -66 & 6.66 & 13.55 & 205 & -82 & H13 & 19.35 & 0.057 & 0.19 & 12.48& 0.40 \\ 
 \hline 
\textbf{1735} & 12.61 & 146 & 11.62 & -74 & 7.19 & 8.25 & 178 & -52 & H16 & 25.46 & 0.097 & 0.52 & 23.06& 0.30 \\ 
 \hline 
\textbf{1884} & 2.89 & 352 & 3.81 & 51 & 3.44 & 2.40 & 42 & 52 & Ď20 & 30.49 & 0.138 & 0.51 & & 0.20 \\ 
 \hline 
\textbf{1927} & 8.56 & 50 & 5.80 & 69 & 2.92 & 4.00 & 74 & 73 & H13 & 8.79 & 0.012 & 0.36 & & 0.50 \\ 
 \hline 
\textbf{2951} & 4.78 & 112 & 11.91 & -71 & 3.68 & 10.00 & 83 & -65 & Ď18 & 12.45 & 0.024 & 0.34 & & 0.50 \\ 
 \hline 
\textbf{3324} & 8.61 & 19 & 3.75 & 84 & 1.17 & 1.00 & 25 & 75 & Ď19 & 9.21 & 0.013 & 0.31 & & 0.50 \\ 
 \hline 
\textbf{4085} & 14.67 & 82 & 23.92 & -76 & 5.44 & 13.80 & 63 & -39 & Ď20 & 37.95 & 0.211 & 0.57 & 32.79& 0.35 \\ 
 \hline 
\textbf{4169} & 10.89 & 81 & 24.97 & -78 & 3.62 & 11.50 & 250 & -72 & Ď18 &29.58 & 0.130 & 0.55 & 28.47& 0.50 \\ 
 \hline 
\textbf{4435} & 4.42 & 353 & 3.84 & -37 & 3.39 & 5.25 & 335 & -13 & Ď20 &28.92 & 0.125 & 0.32 &&  0.35 \\ 
 \hline 
\textbf{4717} & 12.81 & 192 & 9.43 & 10 & 40.78 & 72.50 & 222 & 32 & Ď20 & 35.77 & 0.189 & 0.67 & & 0.50 \\ 
 \hline 
\textbf{4747} & 10.60 & 45 & 22.61 & -81 & 3.43 & 5.90 & 17 & -85 & Ď19 & 5.48 & 0.005 & 0.10 & & 0.30 \\ 
 \hline 
\textbf{5360} & 8.65 & 285 & 4.68 & 83 & 1.91 & 1.50 & 118 & 83 & Ď20 & 13.93 & 0.029 & 0.34 & & 0.50 \\ 
 \hline 
\textbf{8146} & 5.36 & 107 & 17.30 & 76 & 2.97 & 4.90 & 141 & 28 & Ď20 & 50.72 & 0.367 & 0.90 & & 0.25 \\ 
 \hline 
\textbf{10637} & 10.74 & 292 & 6.80 & 66 & 8.27 & 10.80 & 312 & 33 & Ď19 &35.39 & 0.185 & 0.49 & 17.02& 0.50 \\ 
 \hline 
\textbf{12097} & 6.85 & 349 & 2.47 & 81 & 1.92 & 1.50 & 267 & 38 & Ď18 & 51.25 & 0.374 & 0.86 & & 0.50 \\ 
 \hline 
\textbf{12193} & 7.58 & 128 & 9.38 & -76 & 2.95 & 5.00 & 353 & -89 & Ď20 & 14.80 & 0.033 & 0.31 & 31.63& 0.50 \\ 
 \hline 
\textbf{12333} & 31.53 & 158 & 41.66 & -78 & 4.85 & 16.50 & 26 & -35 & Ď19 & 63.17 & 0.549 & 0.68 & & 0.50 \\ 
 \hline 
\textbf{13289} & 463.89 & 144 & 23.63 & 78 & 3.88 & 10.00 & 129 & 41 & Ď19 & 37.80 & 0.210 & 0.76 & 22.17& 0.50 \\ 
 \hline 
\textbf{13297} & 40.13 & 262 & 19.58 & 76 & 3.79 & 11.00 & 242 & 40 & Ď20 & 36.99 & 0.201 & 0.75 & 25.63& 0.50 \\ 
 \hline 
\textbf{13364} & 7.66 & 291 & 13.09 & 72 & 4.46 & 10.00 & 128 & 70 & Ď19 & 37.18 & 0.203 & 0.56 & 38.13& 0.50 \\ 
 \hline 
\textbf{14035} & 153.56 & 216 & 13.56 & -74 & 4.76 & 10.00 & 248 & -63 & Ď20 & 15.46 & 0.036 & 0.59 & & 0.50 \\ 
 \hline 
\textbf{15288} & 7.08 & 7 & 3.32 & 68 & 4.52 & 24.31 & 4.00 & 71 & Ď19 & 17.14 & 0.044 &  0.43 &15.28 & 0.50 \\ 
 \hline 
\textbf{16457} & 14.43 & 4 & 0.31 & 64 & 1.96 & 1.50 & 34 & 22 & Ď20 & 46.49 & 0.311 & 0.69 & & 0.50 \\ 
 \hline 
\textbf{18057} & 8.51 & 296 & 5.57 & -63 & 6.86 & 9.00 & 306 & -31 & Ď18 &2.53 & 0.157 & 0.57 & & 0.50 \\ 
 \hline 
\textbf{18156} & 11.92 & 52 & 21.43 & -78 & 3.65 & 8.50 & 38 & -13 & Ď19 & 65.68 &  0.588 & 0.95 &14.26& 0.50 \\ 
 \hline 
\textbf{20124} & 16.67 & 159 & 26.19 & -75 & 4.98 & 13.00 & 79 & -47 & Ď18 & 42.57 & 0.264 & 0.76 & & 0.50 \\ 
 \hline 
\textbf{20602} & 7.31 & 237 & 12.53 & 74 & 4.05 & 10.00 & 27 & 66 & Ď19 & 39.04 & 0.223 & 0.76 & & 0.50 \\ 
 \hline 
\textbf{24314} & 12.90 & 315 & 24.61 & 76 & 5.09 & 13.50 & 268 & 70 & Ď19 & 14.49 & 0.032 & 0.47 & & 0.50 \\ 
 \hline 
\textbf{31257} & 7.66 & 96 & 21.54 & -78 & 3.78 & 9.50 & 105 & -43 & Ď18 & 35.45 & 0.185 & 0.75 & & 0.50 \\ 
 \hline 
\textbf{36587} & 244.82 & 332 & 15.90 & 70 & 7.58 & 17.00 & 14 & 56 & Ď19 & 23.03 & 0.080 & 0.73 & & 0.50 \\ 
 \hline 
\textbf{66076} & 8.94 & 84 & 19.81 & -77 & 5.31 & 13.00 & 100 & -54 & Ď16 & 23.84 & 0.085 & 0.54 & & 0.50 \\ 
 \hline 
\textbf{70178} & 260.25 & 42 & 18.14 & 76 & 3.51 & 10.00 & 73 & -5 & Ď20 & 82.91 & 0.877 & 0.99 & & 0.50 \\ 
\hline 
\end{tabular}
\end{center}
\caption{\footnotesize{Summary of spin solutions for the asteroids presented in this work, derived from the lowest 2\% of the maximum $\chi^2$ values for each asteroid. The columns are: 
$P$ -- synodic period in hours, obtained from the TESS measurements; 
$\lambda_{\rm T}$ \& $\beta_{\rm T}$ -- TESS pole solutions and uncertainties in ecliptic coordinates; 
 $A$ -- area of the $\chi^2$ region used to derive the pole solutions (white areas on the $\chi^2$ plots). 
 $\lambda_{\rm D}$ \& $\beta_{\rm D}$ -- Pole solutions from the DAMIT database; Ref. -- reference for the DAMIT coordinates, where H13: \cite{2013A&A...551A..67H}, H16: \cite{2016A&A...586A.108H}, H17: \cite{2017A&A...601A.114H}, Ď16: \cite{2016A&A...587A..48D}, Ď18:\cite{2018A&A...617A..57D}, Ď19. \cite{2019A&A...631A...2D}, Ď20: \cite{2020A&A...643A..59D};} '$d$' -- distance between the TESS and DAMIT solutions; 'p' -- probability of the $d$ distances with respect to random sampling ($\propto (1-cos(d))$).  $\Xi$ -- $\chi^2$ value of the TESS solution map at the location of the DAMIT solution;  'd$_{\rm{24}}$' -- distance between closest TESS and C24 solution; 'x' -- $x$ shape parameter from the TESS solution.}
\label{tab:Results} 
\end{table*}

To further test the effectiveness of our method, we applied it to ground-based observations from the DAMIT database, instead of our TESS data. We wanted to see if using a different dataset would still give us similar results to those already published in DAMIT.
From the 44 objects, we selected three asteroids—(22) Kalliope, (94) Aurora, and (1572) Posnania— that had at least three, dense light curve measurements in DAMIT, with spacial distributions similar to those obtained from TESS.

Using the DAMIT data, we followed the same steps as before and we created the same $\chi^2$ maps to highlight the most likely solutions (Figs.~4--6 in Appendix \ref{App:figs}). In the given plots, the white pixels once again highlight the most likely pole solutions, representing the areas where the $\chi^2$ values are the lowest 2\% of the maximum $\chi^2$ value.

The results from the DAMIT data showed some similarities to those obtained from the TESS data, with the closest agreement found for (22) Kalliope and (1572) Posnania. 
However, when we use light curves that are similar in number to those from TESS (at least three dense light curves) but obtained from other sources (ground-based observations as listed in DAMIT) without any additional constraints on the shape or amplitude variations outside the observed epochs, the pole orientations are similar to the original spin solutions stored in DAMIT; however, they still do not completely align with them.

\begin{figure*}[ht!]
    \centering
    \includegraphics[width=0.8\textwidth]{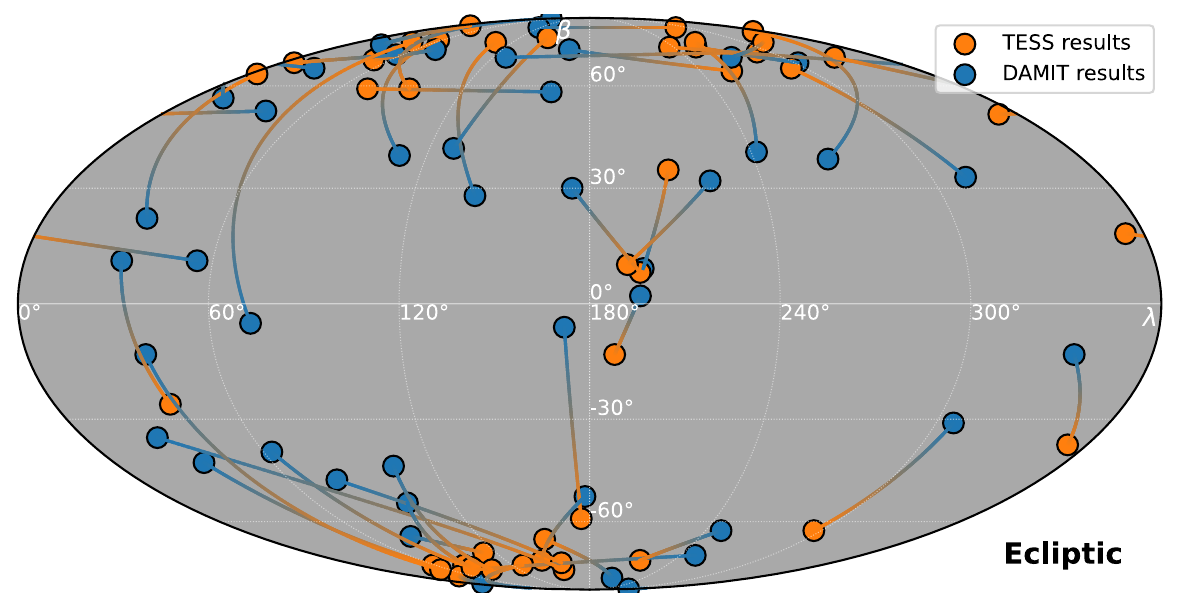}
    \caption{Spin axis solution comparison for the selected 44 objects: The blue and orange-filled circles represent the DAMIT and our results, respectively.}
    \label{fig:HealFigComparisonOfCoordinates}
\end{figure*}

In Fig. \ref{fig:HealFigComparisonOfCoordinates} we present our pole-orientation results for all the 44 selected asteroids, along with the solutions given in DAMIT. For each asteroid, a pair of orange and blue dots represent the pole solutions derived from TESS measurements and the DAMIT database, respectively (see also Table~\ref{tab:Results}).

\begin{figure*}
    \centering
    \includegraphics[width=1\textwidth]{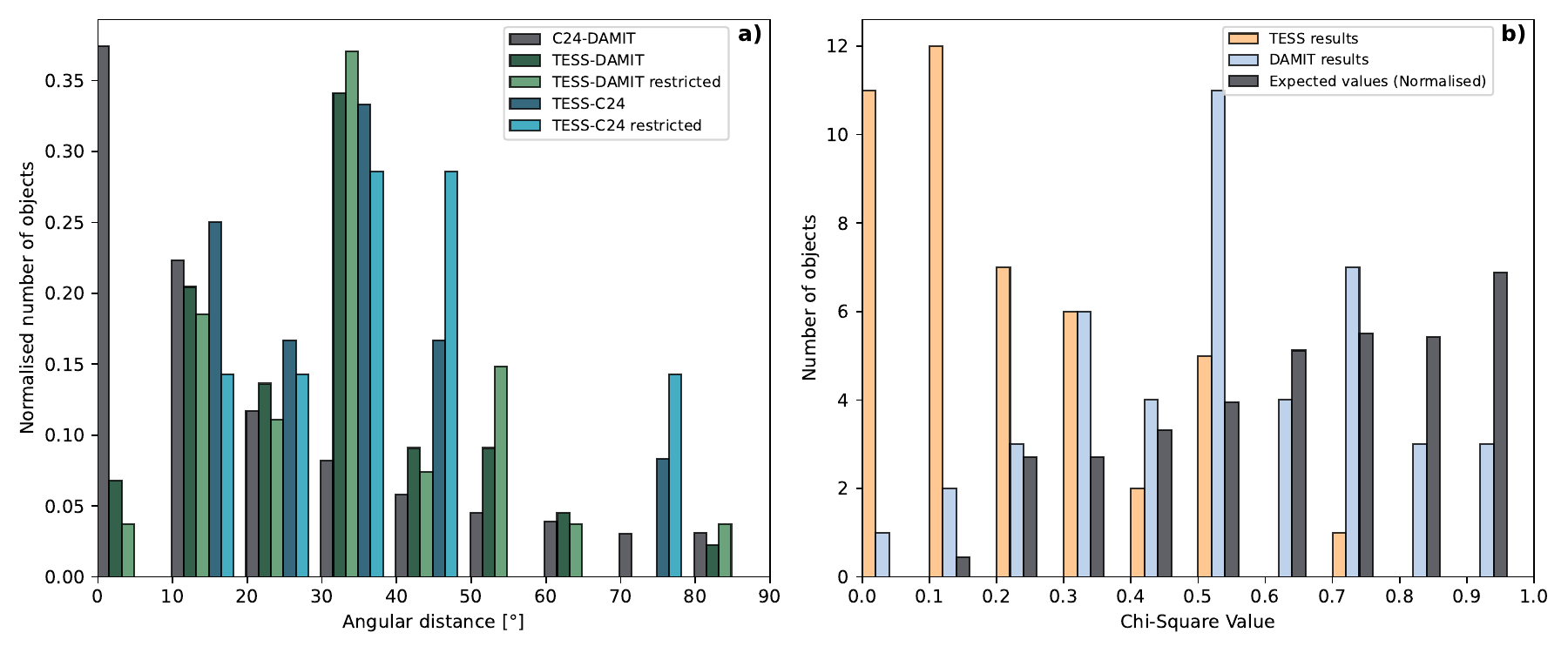}
    \caption{Summary figures of TESS, DAMIT, and C24 solution comparisons. a) Distribution of angular distances between our TESS solutions and those from DAMIT and C24. We compare the DAMIT-C24, TESS-DAMIT and TESS-C24 differences, as well as the restricted cases where only asteroids observed in at least four sectors are considered. b) Distribution of $\chi^2$  values obtained from $\chi^2$ maps basing on our results, using TESS light curve data. The $\chi^2$ samples were taken at the location of the original DAMIT solutions (blue) and our derived TESS pole solutions (orange). The grey bars show the distribution expected from a random sampling of the same maps.}
\label{fig:FourPanelComparisonOfCoordinatesAndDistances}
\end{figure*}

In addition to DAMIT, we also compared our results with those in \citet[][C24]{Cellino2024}, where the common asteroids are (1381) Danubia, (1572) Posnania, (1735) ITA, (4085) Weir, (4169) Celsius, 10637 Heimlich, (12193) 1979 EL, (13289) 1998 QK75, (13297) 1998 RX, (13364) 1998 UK20, (15288) 1991 RN27 and (18156) Kamisaibara. For these 12 targets, our $\chi^2$ plots also mark the C24 solutions (see, for example, Fig.~6. in Appendix~\ref{App:figs}). 

The distances between the C24 and TESS pole solutions for these 12 asteroids are also presented in Table~\ref{tab:Results}. In most cases (9 out of 12), the distance between the C24 and our TESS results is lower than the distance between our results and the DAMIT solution.

These distances are also presented in Fig.~\ref{fig:FourPanelComparisonOfCoordinatesAndDistances}a, which shows how the angular differences between the pole solutions from all three datasets are distributed, and how the TESS results compare to both C24 and DAMIT, as well as the difference between the C24 and the DAMIT solutions.

\section{Discussion}
\label{sec:summary}

The comparison between our TESS-derived pole solutions and those from the DAMIT database shows that, in general, there is a relatively large angular distance between the two sets of results, with an average separation of $\sim 33^{\circ}$. A similar discrepancy is observed when comparing our TESS results, which assume triaxial ellipsoid shapes, with the C24 pole solutions, which are based on the same shape assumption for the asteroids (see Table~\ref{tab:Results}). For the twelve objects common to both datasets, the average difference between our solutions and those from C24 is approximately $\sim \mathbf{25}^{\circ}$. 

These results highlight the limitations of low-degree-of-freedom models in asteroid light curve inversion. While such models allow for the efficient analysis of large survey datasets, our findings demonstrate that TESS data alone are often insufficient for reliably reconstructing spin axis orientations. In many cases, additional observations from other sources are necessary to achieve more precise solutions.

Additionally, we examined the differences between the C24 and DAMIT pole solutions. By identifying the common objects between these datasets, we found a sample of approximately 3800 asteroids. The average angular distance between the C24 and DAMIT pole solutions is $\sim 24^{\circ}$, which is notably smaller than the differences observed between our TESS-based solutions and either dataset (see Fig.~\ref{fig:FourPanelComparisonOfCoordinatesAndDistances}a).

It is, however, typically the case that both the TESS and DAMIT pole solutions -- as well as the C24 solutions, if available -- are located in the low $\chi^2$ regions of the TESS solutions, as demonstrated by the $\chi^2$ maps, and also in Fig.~\ref{fig:FourPanelComparisonOfCoordinatesAndDistances}b. The limited number of observed orbital phases, typically a maximum of 3-4 in the presently available TESS data, does not allow a better mapping of these low $\chi^2$ regions. In many instances, we find that the pole solutions from DAMIT are bracketed by two TESS-derived pole solutions in low $\chi^2$ pole solution areas.

We also tested our method by excluding objects with only three measurements, leaving only those with four or more sector measurements that were relatively far from each other. As a result, we identified 27 common objects for comparison in DAMIT, instead of the 44, and 7 in C24. However, when we compared our results with those of DAMIT and C24, the average angular distances did not show any improvement. They show an average of $\sim 35^{\circ}$ and $\sim 39^{\circ}$ angular distances when comparing our results to the DAMIT and C24 solutions, respectively (see Fig.~\ref{fig:FourPanelComparisonOfCoordinatesAndDistances}a). Nevertheless, in most cases, we were able to confidently identify regions with high $\chi^2$ values, where the pole solution is unlikely to be.

This analysis shows that excluding objects with less than four sector measurements does not improve the results. Even considering the targets with the largest number of available TESS sectors (e.g. six sectors for (4717) Kaneko) provides an angular distance between the TESS and DAMIT solutions, similar to the average value ($35.77^{\circ}$).
Since the goal of this study was to assess the applicability of shape modelling for main belt asteroids using a simple shape approximation and limited directional photometry from TESS, we did not apply further restrictions on the dataset.

Given the number of sector measurements per asteroid and the sampling characteristics of TESS, this method reaches its practical limits. While TESS provides precise measurements of the amplitude and period within individual sectors, it lacks information on how these properties change over time. In this respect, a dataset that covers multiple points along an asteroid's orbit - albeit with less detail per observation - provides better constraints on shape and spin axis orientation than a dataset that samples fewer points but does so with more accuracy \citep{2016A&A...595A...1G,2018PASP..130f4505T}.

Our statistical analysis demonstrates that, in most cases, we are able to reliably identify the regions where the rotation axis is unlikely to be located. While the exact pole orientations do not always converge with the previously published DAMIT solutions, our results show a clear overlap in the low $\chi˘{2}$ regions, indicating that our method captures the general trends of the spin-axis distribution. This suggests that even with the limited viewing geometry provided by TESS, our approach is capable of constraining the possible pole orientations with reasonable accuracy,  for around half of the studied objects. The discrepancies between our solutions and those from DAMIT are primarily due to the inherent limitations of the TESS dataset, particularly the restricted number of observation sectors and viewing angles available for each asteroid. However, our results reaffirm that, despite these constraints, TESS-based modelling can still provide valuable insights into the rotational properties of asteroids, especially when combined with complementary datasets in future studies.

\facilities{TESS \citep{2015JATIS...1a4003R}
} 
\software{FITSH \citep{pal2012}, EPHEMD}

\vspace{5mm}
% \begin{acknowledgements}
\section*{Acknowledgements}
The research leading to these results has received funding from the K-138962 and TKP2021-NKTA-64 grants of the National Research, Development and Innovation Office (NKFIH, Hungary).
Funding for Minor Planet Center (MPC) data and operations comes from a NASA PDCO grant (80NSSC22M0024), administered via a University of Maryland - SAO subaward (106075-Z6415201). The MPC's computing equipment is funded in part by the above award, and in part by funding from the Tamkin Foundation. 
Some of the results in this paper have been derived using the healpy and HEALPix package.
We are indebted to our reviewer for the useful comments and suggestions. 

% \end{acknowledgements}

\newpage
\bibliography{tedr2sp}

\begin{thebibliography}{}
\expandafter\ifx\csname natexlab\endcsname\relax\def\natexlab#1{#1}\fi
\providecommand{\url}[1]{\href{#1}{#1}}
\providecommand{\dodoi}[1]{doi:~\href{http://doi.org/#1}{\nolinkurl{#1}}}
\providecommand{\doeprint}[1]{\href{http://ascl.net/#1}{\nolinkurl{http://ascl.net/#1}}}
\providecommand{\doarXiv}[1]{\href{https://arxiv.org/abs/#1}{\nolinkurl{https://arxiv.org/abs/#1}}}

\bibitem[{{Cellino} {et~al.}(2024){Cellino}, {Tanga}, {Muinonen}, \&
  {Mignard}}]{Cellino2024}
{Cellino}, A., {Tanga}, P., {Muinonen}, K., \& {Mignard}, F. 2024, \aap, 687,
  A277, \dodoi{10.1051/0004-6361/202449297}

\bibitem[{{Farkas-Tak{\'a}cs} {et~al.}(2017){Farkas-Tak{\'a}cs}, {Kiss},
  {P{\'a}l}, {Moln{\'a}r}, {Szab{\'o}}, {Hanyecz}, {S{\'a}rneczky},
  {Szab{\'o}}, {Marton}, {Mommert}, {Szak{\'a}ts}, {M{\"u}ller}, \&
  {Kiss}}]{farkas2017}
{Farkas-Tak{\'a}cs}, A., {Kiss}, C., {P{\'a}l}, A., {et~al.} 2017, \aj, 154,
  119, \dodoi{10.3847/1538-3881/aa8365}

\bibitem[{{Gaia Collaboration} {et~al.}(2016){Gaia Collaboration}, {Prusti},
  {de Bruijne}, {Brown}, {Vallenari}, {Babusiaux}, {Bailer-Jones}, {Bastian},
  {Biermann}, {Evans}, {Eyer}, {Jansen}, {Jordi}, {Klioner}, {Lammers},
  {Lindegren}, {Luri}, {Mignard}, {Milligan}, {Panem}, {Poinsignon},
  {Pourbaix}, {Randich}, {Sarri}, {Sartoretti}, {Siddiqui}, {Soubiran},
  {Valette}, {van Leeuwen}, {Walton}, {Aerts}, {Arenou}, {Cropper}, {Drimmel},
  {H{\o}g}, {Katz}, {Lattanzi}, {O'Mullane}, {Grebel}, {Holland}, {Huc},
  {Passot}, {Bramante}, {Cacciari}, {Casta{\~n}eda}, {Chaoul}, {Cheek}, {De
  Angeli}, {Fabricius}, {Guerra}, {Hern{\'a}ndez}, {Jean-Antoine-Piccolo},
  {Masana}, {Messineo}, {Mowlavi}, {Nienartowicz}, {Ord{\'o}{\~n}ez-Blanco},
  {Panuzzo}, {Portell}, {Richards}, {Riello}, {Seabroke}, {Tanga},
  {Th{\'e}venin}, {Torra}, {Els}, {Gracia-Abril}, {Comoretto},
  {Garcia-Reinaldos}, {Lock}, {Mercier}, {Altmann}, {Andrae}, {Astraatmadja},
  {Bellas-Velidis}, {Benson}, {Berthier}, {Blomme}, {Busso}, {Carry},
  {Cellino}, {Clementini}, {Cowell}, {Creevey}, {Cuypers}, {Davidson}, {De
  Ridder}, {de Torres}, {Delchambre}, {Dell'Oro}, {Ducourant}, {Fr{\'e}mat},
  {Garc{\'\i}a-Torres}, {Gosset}, {Halbwachs}, {Hambly}, {Harrison}, {Hauser},
  {Hestroffer}, {Hodgkin}, {Huckle}, {Hutton}, {Jasniewicz}, {Jordan},
  {Kontizas}, {Korn}, {Lanzafame}, {Manteiga}, {Moitinho}, {Muinonen},
  {Osinde}, {Pancino}, {Pauwels}, {Petit}, {Recio-Blanco}, {Robin}, {Sarro},
  {Siopis}, {Smith}, {Smith}, {Sozzetti}, {Thuillot}, {van Reeven}, {Viala},
  {Abbas}, {Abreu Aramburu}, {Accart}, {Aguado}, {Allan}, {Allasia},
  {Altavilla}, {{\'A}lvarez}, {Alves}, {Anderson}, {Andrei}, {Anglada Varela},
  {Antiche}, {Antoja}, {Ant{\'o}n}, {Arcay}, {Atzei}, {Ayache}, {Bach},
  {Baker}, {Balaguer-N{\'u}{\~n}ez}, {Barache}, {Barata}, {Barbier}, {Barblan},
  {Baroni}, {Barrado y Navascu{\'e}s}, {Barros}, {Barstow}, {Becciani},
  {Bellazzini}, {Bellei}, {Bello Garc{\'\i}a}, {Belokurov}, {Bendjoya},
  {Berihuete}, {Bianchi}, {Bienaym{\'e}}, {Billebaud}, {Blagorodnova},
  {Blanco-Cuaresma}, {Boch}, {Bombrun}, {Borrachero}, {Bouquillon}, {Bourda},
  {Bouy}, {Bragaglia}, {Breddels}, {Brouillet}, {Br{\"u}semeister},
  {Bucciarelli}, {Budnik}, {Burgess}, {Burgon}, {Burlacu}, {Busonero}, {Buzzi},
  {Caffau}, {Cambras}, {Campbell}, {Cancelliere}, {Cantat-Gaudin}, {Carlucci},
  {Carrasco}, {Castellani}, {Charlot}, {Charnas}, {Charvet}, {Chassat},
  {Chiavassa}, {Clotet}, {Cocozza}, {Collins}, {Collins}, \&
  {Costigan}}]{2016A&A...595A...1G}
{Gaia Collaboration}, {Prusti}, T., {de Bruijne}, J.~H.~J., {et~al.} 2016,
  \aap, 595, A1, \dodoi{10.1051/0004-6361/201629272}

\bibitem[{{G{\'o}rski} {et~al.}(2005){G{\'o}rski}, {Hivon}, {Banday},
  {Wandelt}, {Hansen}, {Reinecke}, \& {Bartelmann}}]{2005ApJ...622..759G}
{G{\'o}rski}, K.~M., {Hivon}, E., {Banday}, A.~J., {et~al.} 2005, \apj, 622,
  759, \dodoi{10.1086/427976}

\bibitem[{{Hanu{\v{s}}} {et~al.}(2023){Hanu{\v{s}}}, {Vokrouhlick{\'y}},
  {Nesvorn{\'y}}, {{\v{D}}urech}, {Stephens}, {Benishek}, {Oey}, \&
  {Pokorn{\'y}}}]{Hanus2023}
{Hanu{\v{s}}}, J., {Vokrouhlick{\'y}}, D., {Nesvorn{\'y}}, D., {et~al.} 2023,
  \aap, 679, A56, \dodoi{10.1051/0004-6361/202346022}

\bibitem[{{Hanu{\v{s}}} {et~al.}(2013){Hanu{\v{s}}}, {{\v{D}}urech},
  {Bro{\v{z}}}, {Marciniak}, {Warner}, {Pilcher}, {Stephens}, {Behrend},
  {Carry}, {{\v{C}}apek}, {Antonini}, {Audejean}, {Augustesen}, {Barbotin},
  {Baudouin}, {Bayol}, {Bernasconi}, {Borczyk}, {Bosch}, {Brochard},
  {Brunetto}, {Casulli}, {Cazenave}, {Charbonnel}, {Christophe}, {Colas},
  {Coloma}, {Conjat}, {Cooney}, {Correira}, {Cotrez}, {Coupier}, {Crippa},
  {Cristofanelli}, {Dalmas}, {Danavaro}, {Demeautis}, {Droege}, {Durkee},
  {Esseiva}, {Esteban}, {Fagas}, {Farroni}, {Fauvaud}, {Fauvaud}, {Del Freo},
  {Garcia}, {Geier}, {Godon}, {Grangeon}, {Hamanowa}, {Hamanowa}, {Heck},
  {Hellmich}, {Higgins}, {Hirsch}, {Husarik}, {Itkonen}, {Jade},
  {Kami{\'n}ski}, {Kankiewicz}, {Klotz}, {Koff}, {Kryszczy{\'n}ska},
  {Kwiatkowski}, {Laffont}, {Leroy}, {Lecacheux}, {Leonie}, {Leyrat},
  {Manzini}, {Martin}, {Masi}, {Matter}, {Micha{\l}owski}, {Micha{\l}owski},
  {Micha{\l}owski}, {Michelet}, {Michelsen}, {Morelle}, {Mottola}, {Naves},
  {Nomen}, {Oey}, {Og{\l}oza}, {Oksanen}, {Oszkiewicz},
  {P{\"a}{\"a}kk{\"o}nen}, {Paiella}, {Pallares}, {Paulo}, {Pavic}, {Payet},
  {Poli{\'n}ska}, {Polishook}, {Poncy}, {Revaz}, {Rinner}, {Rocca}, {Roche},
  {Romeuf}, {Roy}, {Saguin}, {Salom}, {Sanchez}, {Santacana}, {Santana-Ros},
  {Sareyan}, {Sobkowiak}, {Sposetti}, {Starkey}, {Stoss}, {Strajnic}, {Teng},
  {Tr{\'e}gon}, {Vagnozzi}, {Velichko}, {Waelchli}, {Wagrez}, \&
  {W{\"u}cher}}]{2013A&A...551A..67H}
{Hanu{\v{s}}}, J., {{\v{D}}urech}, J., {Bro{\v{z}}}, M., {et~al.} 2013, \aap,
  551, A67, \dodoi{10.1051/0004-6361/201220701}

\bibitem[{{Hanu{\v{s}}} {et~al.}(2016){Hanu{\v{s}}}, {{\v{D}}urech},
  {Oszkiewicz}, {Behrend}, {Carry}, {Delbo}, {Adam}, {Afonina}, {Anquetin},
  {Antonini}, {Arnold}, {Audejean}, {Aurard}, {Bachschmidt}, {Baduel},
  {Barbotin}, {Barroy}, {Baudouin}, {Berard}, {Berger}, {Bernasconi}, {Bosch},
  {Bouley}, {Bozhinova}, {Brinsfield}, {Brunetto}, {Canaud}, {Caron},
  {Carrier}, {Casalnuovo}, {Casulli}, {Cerda}, {Chalamet}, {Charbonnel},
  {Chinaglia}, {Cikota}, {Colas}, {Coliac}, {Collet}, {Coloma}, {Conjat},
  {Conseil}, {Costa}, {Crippa}, {Cristofanelli}, {Damerdji}, {Deback{\`e}re},
  {Decock}, {D{\'e}hais}, {D{\'e}l{\'e}age}, {Delmelle}, {Demeautis},
  {Dr{\'o}{\.z}d{\.z}}, {Dubos}, {Dulcamara}, {Dumont}, {Durkee}, {Dymock},
  {Escalante del Valle}, {Esseiva}, {Esseiva}, {Esteban}, {Fauchez},
  {Fauerbach}, {Fauvaud}, {Fauvaud}, {Forn{\'e}}, {Fournel}, {Fradet},
  {Garlitz}, {Gerteis}, {Gillier}, {Gillon}, {Giraud}, {Godard}, {Goncalves},
  {Hamanowa}, {Hamanowa}, {Hay}, {Hellmich}, {Heterier}, {Higgins}, {Hirsch},
  {Hodosan}, {Hren}, {Hygate}, {Innocent}, {Jacquinot}, {Jawahar}, {Jehin},
  {Jerosimic}, {Klotz}, {Koff}, {Korlevic}, {Kosturkiewicz}, {Krafft},
  {Krugly}, {Kugel}, {Labrevoir}, {Lecacheux}, {Lehk{\'y}}, {Leroy},
  {Lesquerbault}, {Lopez-Gonzales}, {Lutz}, {Mallecot}, {Manfroid}, {Manzini},
  {Marciniak}, {Martin}, {Modave}, {Montaigut}, {Montier}, {Morelle}, {Morton},
  {Mottola}, {Naves}, {Nomen}, {Oey}, {Og{\l}oza}, {Paiella}, {Pallares},
  {Peyrot}, {Pilcher}, {Pirenne}, {Piron}, {Poli{\'n}ska}, {Polotto}, {Poncy},
  {Previt}, {Reignier}, {Renauld}, {Ricci}, {Richard}, {Rinner}, {Risoldi},
  {Robilliard}, {Romeuf}, {Rousseau}, {Roy}, {Ruthroff}, {Salom}, {Salvador},
  {Sanchez}, {Santana-Ros}, {Scholz}, {S{\'e}n{\'e}}, {Skiff}, {Sobkowiak},
  {Sogorb}, {Sold{\'a}n}, {Spiridakis}, {Splanska}, {Sposetti}, {Starkey},
  {Stephens}, {Stiepen}, {Stoss}, {Strajnic}, {Teng}, {Tumolo}, {Vagnozzi},
  {Vanoutryve}, {Vugnon}, {Warner}, {Waucomont}, {Wertz}, {Winiarski}, \&
  {Wolf}}]{2016A&A...586A.108H}
{Hanu{\v{s}}}, J., {{\v{D}}urech}, J., {Oszkiewicz}, D.~A., {et~al.} 2016,
  \aap, 586, A108, \dodoi{10.1051/0004-6361/201527441}

\bibitem[{{Hanu{\v{s}}} {et~al.}(2017){Hanu{\v{s}}}, {Viikinkoski}, {Marchis},
  {{\v{D}}urech}, {Kaasalainen}, {Delbo'}, {Herald}, {Frappa}, {Hayamizu},
  {Kerr}, {Preston}, {Timerson}, {Dunham}, \& {Talbot}}]{2017A&A...601A.114H}
{Hanu{\v{s}}}, J., {Viikinkoski}, M., {Marchis}, F., {et~al.} 2017, \aap, 601,
  A114, \dodoi{10.1051/0004-6361/201629956}

\bibitem[{{Holman} {et~al.}(2019){Holman}, {Payne}, \& {P{\'a}l}}]{holman2019}
{Holman}, M.~J., {Payne}, M.~J., \& {P{\'a}l}, A. 2019, Research Notes of the
  American Astronomical Society, 3, 160, \dodoi{10.3847/2515-5172/ab4ea6}

\bibitem[{{Howell} {et~al.}(2014){Howell}, {Sobeck}, {Haas}, {Still},
  {Barclay}, {Mullally}, {Troeltzsch}, {Aigrain}, {Bryson}, {Caldwell},
  {Chaplin}, {Cochran}, {Huber}, {Marcy}, {Miglio}, {Najita}, {Smith},
  {Twicken}, \& {Fortney}}]{howell2014}
{Howell}, S.~B., {Sobeck}, C., {Haas}, M., {et~al.} 2014, \pasp, 126, 398,
  \dodoi{10.1086/676406}

\bibitem[{Kiss {et~al.}(2020)Kiss, Molnár, Pál, \&
  Howell}]{10.1088/2514-3433/ab9823ch5}
Kiss, C., Molnár, L., Pál, A., \& Howell, S.~B. 2020, in The NASA Kepler
  Mission, 2514-3433 (IOP Publishing), 5--1 to 5--14,
  \dodoi{10.1088/2514-3433/ab9823ch5}

\bibitem[{{Kiss} {et~al.}(2016){Kiss}, {P{\'a}l}, {Farkas-Tak{\'a}cs},
  {Szab{\'o}}, {Szab{\'o}}, {Kiss}, {Moln{\'a}r}, {S{\'a}rneczky},
  {M{\"u}ller}, {Mommert}, \& {Stansberry}}]{kiss2016}
{Kiss}, C., {P{\'a}l}, A., {Farkas-Tak{\'a}cs}, A.~I., {et~al.} 2016, \mnras,
  457, 2908, \dodoi{10.1093/mnras/stw081}

\bibitem[{{Kiss} {et~al.}(2017){Kiss}, {Marton}, {Farkas-Tak{\'a}cs},
  {Stansberry}, {M{\"u}ller}, {Vink{\'o}}, {Balog}, {Ortiz}, \&
  {P{\'a}l}}]{kiss2017}
{Kiss}, C., {Marton}, G., {Farkas-Tak{\'a}cs}, A., {et~al.} 2017, \apjl, 838,
  L1, \dodoi{10.3847/2041-8213/aa6484}

\bibitem[{{Magnusson}(1986)}]{1986Icar...68....1M}
{Magnusson}, P. 1986, \icarus, 68, 1, \dodoi{10.1016/0019-1035(86)90072-2}

\bibitem[{{McNeill} {et~al.}(2019){McNeill}, {Mommert}, {Trilling}, {Llama}, \&
  {Skiff}}]{mcneill2019}
{McNeill}, A., {Mommert}, M., {Trilling}, D.~E., {Llama}, J., \& {Skiff}, B.
  2019, \apjs, 245, 29, \dodoi{10.3847/1538-4365/ab5223}

\bibitem[{{Moln{\'a}r} {et~al.}(2018){Moln{\'a}r}, {P{\'a}l}, {S{\'a}rneczky},
  {Szab{\'o}}, {Vink{\'o}}, {Szab{\'o}}, {Kiss}, {Hanyecz}, {Marton}, \&
  {Kiss}}]{molnar2018}
{Moln{\'a}r}, L., {P{\'a}l}, A., {S{\'a}rneczky}, K., {et~al.} 2018, \apjs,
  234, 37, \dodoi{10.3847/1538-4365/aaa1a1}

\bibitem[{{Muinonen} \& {Lumme}(2015)}]{ML2015}
{Muinonen}, K., \& {Lumme}, K. 2015, \aap, 584, A23,
  \dodoi{10.1051/0004-6361/201526456}

\bibitem[{{Muinonen} {et~al.}(2015){Muinonen}, {Wilkman}, {Cellino}, {Wang}, \&
  {Wang}}]{Muinonen2015}
{Muinonen}, K., {Wilkman}, O., {Cellino}, A., {Wang}, X., \& {Wang}, Y. 2015,
  \planss, 118, 227, \dodoi{10.1016/j.pss.2015.09.005}

\bibitem[{{P{\'a}l}(2012)}]{pal2012}
{P{\'a}l}, A. 2012, \mnras, 421, 1825, \dodoi{10.1111/j.1365-2966.2011.19813.x}

\bibitem[{{P{\'a}l} {et~al.}(2016){P{\'a}l}, {Kiss}, {M{\"u}ller},
  {Moln{\'a}r}, {Szab{\'o}}, {Szab{\'o}}, {S{\'a}rneczky}, \& {Kiss}}]{pal2016}
{P{\'a}l}, A., {Kiss}, C., {M{\"u}ller}, T.~G., {et~al.} 2016, \aj, 151, 117,
  \dodoi{10.3847/0004-6256/151/5/117}

\bibitem[{{P{\'a}l} {et~al.}(2018){P{\'a}l}, {Moln{\'a}r}, \& {Kiss}}]{pal2018}
{P{\'a}l}, A., {Moln{\'a}r}, L., \& {Kiss}, C. 2018, \pasp, 130, 114503,
  \dodoi{10.1088/1538-3873/aae2aa}

\bibitem[{{P{\'a}l} {et~al.}(2015){P{\'a}l}, {Szab{\'o}}, {Szab{\'o}}, {Kiss},
  {Moln{\'a}r}, {S{\'a}rneczky}, \& {Kiss}}]{pal2015}
{P{\'a}l}, A., {Szab{\'o}}, R., {Szab{\'o}}, G.~M., {et~al.} 2015, \apjl, 804,
  L45, \dodoi{10.1088/2041-8205/804/2/L45}

\bibitem[{{P{\'a}l} {et~al.}(2020){P{\'a}l}, {Szak{\'a}ts}, {Kiss}, {B{\'o}di},
  {Bogn{\'a}r}, {Kalup}, {Kiss}, {Marton}, {Moln{\'a}r}, {Plachy},
  {S{\'a}rneczky}, {Szab{\'o}}, \& {Szab{\'o}}}]{pal2020}
{P{\'a}l}, A., {Szak{\'a}ts}, R., {Kiss}, C., {et~al.} 2020, \apjs, 247, 26,
  \dodoi{10.3847/1538-4365/ab64f0}

\bibitem[{{Payne} {et~al.}(2019){Payne}, {Holman}, \& {P{\'a}l}}]{payne2019}
{Payne}, M.~J., {Holman}, M.~J., \& {P{\'a}l}, A. 2019, Research Notes of the
  American Astronomical Society, 3, 172, \dodoi{10.3847/2515-5172/ab5641}

\bibitem[{{Rice} \& {Laughlin}(2020)}]{rice2020}
{Rice}, M., \& {Laughlin}, G. 2020, PSJ, 1, 81, \dodoi{10.3847/PSJ/abc42c}

\bibitem[{{Ricker} {et~al.}(2015){Ricker}, {Winn}, {Vanderspek}, {Latham},
  {Bakos}, {Bean}, {Berta-Thompson}, {Brown}, {Buchhave}, {Butler}, {Butler},
  {Chaplin}, {Charbonneau}, {Christensen-Dalsgaard}, {Clampin}, {Deming},
  {Doty}, {De Lee}, {Dressing}, {Dunham}, {Endl}, {Fressin}, {Ge}, {Henning},
  {Holman}, {Howard}, {Ida}, {Jenkins}, {Jernigan}, {Johnson}, {Kaltenegger},
  {Kawai}, {Kjeldsen}, {Laughlin}, {Levine}, {Lin}, {Lissauer}, {MacQueen},
  {Marcy}, {McCullough}, {Morton}, {Narita}, {Paegert}, {Palle}, {Pepe},
  {Pepper}, {Quirrenbach}, {Rinehart}, {Sasselov}, {Sato}, {Seager},
  {Sozzetti}, {Stassun}, {Sullivan}, {Szentgyorgyi}, {Torres}, {Udry}, \&
  {Villasenor}}]{2015JATIS...1a4003R}
{Ricker}, G.~R., {Winn}, J.~N., {Vanderspek}, R., {et~al.} 2015, Journal of
  Astronomical Telescopes, Instruments, and Systems, 1, 014003,
  \dodoi{10.1117/1.JATIS.1.1.014003}

\bibitem[{{Santana-Ros} {et~al.}(2015){Santana-Ros}, {Bartczak},
  {Micha{\l}owski}, {Tanga}, \& {Cellino}}]{SantanaRos2015}
{Santana-Ros}, T., {Bartczak}, P., {Micha{\l}owski}, T., {Tanga}, P., \&
  {Cellino}, A. 2015, \mnras, 450, 333, \dodoi{10.1093/mnras/stv631}

\bibitem[{{Szab{\'o}} {et~al.}(2016){Szab{\'o}}, {P{\'a}l}, {S{\'a}rneczky},
  {Szab{\'o}}, {Moln{\'a}r}, {Kiss}, {Hanyecz}, {Plachy}, \&
  {Kiss}}]{szabo2016}
{Szab{\'o}}, R., {P{\'a}l}, A., {S{\'a}rneczky}, K., {et~al.} 2016, \aap, 596,
  A40, \dodoi{10.1051/0004-6361/201629059}

\bibitem[{{Szak{\'a}ts} {et~al.}(2017){Szak{\'a}ts}, {Kiss}, {Marton},
  {Varga-Vereb{\'e}lyi}, {M{\"u}ller}, \& {P{\'a}l}}]{szakats2017}
{Szak{\'a}ts}, R., {Kiss}, C., {Marton}, G., {et~al.} 2017, in European
  Planetary Science Congress, EPSC2017--223

\bibitem[{{Szak{\'a}ts} {et~al.}(2020){Szak{\'a}ts}, {M{\"u}ller},
  {Al{\'\i}-Lagoa}, {Marton}, {Farkas-Tak{\'a}cs}, {B{\'a}nyai}, \&
  {Kiss}}]{szakats2020}
{Szak{\'a}ts}, R., {M{\"u}ller}, T., {Al{\'\i}-Lagoa}, V., {et~al.} 2020, \aap,
  635, A54, \dodoi{10.1051/0004-6361/201936142}

\bibitem[{{Tenenbaum} \& {Jenkins}(2018)}]{tenenbaum2018}
{Tenenbaum}, P., \& {Jenkins}, J.~M. 2018, {TESS Science Data Products
  Description Document: EXP-TESS-ARC-ICD-0014, Rev. D}

\bibitem[{{Tonry} {et~al.}(2018){Tonry}, {Denneau}, {Heinze}, {Stalder},
  {Smith}, {Smartt}, {Stubbs}, {Weiland}, \& {Rest}}]{2018PASP..130f4505T}
{Tonry}, J.~L., {Denneau}, L., {Heinze}, A.~N., {et~al.} 2018, \pasp, 130,
  064505, \dodoi{10.1088/1538-3873/aabadf}

\bibitem[{{{\v{D}}urech} {et~al.}(2015){{\v{D}}urech}, {Carry}, {Delbo},
  {Kaasalainen}, \& {Viikinkoski}}]{2015aste.book..183D}
{{\v{D}}urech}, J., {Carry}, B., {Delbo}, M., {Kaasalainen}, M., \&
  {Viikinkoski}, M. 2015, in Asteroids IV, ed. P.~{Michel}, F.~E. {DeMeo}, \&
  W.~F. {Bottke}, 183--202, \dodoi{10.2458/azu_uapress_9780816532131-ch010}

\bibitem[{{{\v{D}}urech} \& {Hanu{\v{s}}}(2023)}]{Durech2023}
{{\v{D}}urech}, J., \& {Hanu{\v{s}}}, J. 2023, \aap, 675, A24,
  \dodoi{10.1051/0004-6361/202345889}

\bibitem[{{{\v{D}}urech} {et~al.}(2018){{\v{D}}urech}, {Hanu{\v{s}}}, \&
  {Al{\'\i}-Lagoa}}]{2018A&A...617A..57D}
{{\v{D}}urech}, J., {Hanu{\v{s}}}, J., \& {Al{\'\i}-Lagoa}, V. 2018, \aap, 617,
  A57, \dodoi{10.1051/0004-6361/201833437}

\bibitem[{{{\v{D}}urech} {et~al.}(2016){{\v{D}}urech}, {Hanu{\v{s}}},
  {Oszkiewicz}, \& {Van{\v{c}}o}}]{2016A&A...587A..48D}
{{\v{D}}urech}, J., {Hanu{\v{s}}}, J., {Oszkiewicz}, D., \& {Van{\v{c}}o}, R.
  2016, \aap, 587, A48, \dodoi{10.1051/0004-6361/201527573}

\bibitem[{{{\v{D}}urech} {et~al.}(2019){{\v{D}}urech}, {Hanu{\v{s}}}, \&
  {Van{\v{c}}o}}]{2019A&A...631A...2D}
{{\v{D}}urech}, J., {Hanu{\v{s}}}, J., \& {Van{\v{c}}o}, R. 2019, \aap, 631,
  A2, \dodoi{10.1051/0004-6361/201936341}

\bibitem[{{{\v{D}}urech} {et~al.}(2010){{\v{D}}urech}, {Sidorin}, \&
  {Kaasalainen}}]{durech2010}
{{\v{D}}urech}, J., {Sidorin}, V., \& {Kaasalainen}, M. 2010, \aap, 513, A46,
  \dodoi{10.1051/0004-6361/200912693}

\bibitem[{{{\v{D}}urech} {et~al.}(2020){{\v{D}}urech}, {Tonry}, {Erasmus},
  {Denneau}, {Heinze}, {Flewelling}, \& {Van{\v{c}}o}}]{2020A&A...643A..59D}
{{\v{D}}urech}, J., {Tonry}, J., {Erasmus}, N., {et~al.} 2020, \aap, 643, A59,
  \dodoi{10.1051/0004-6361/202037729}

\bibitem[{{Villase\~{n}or} {et~al.}(2019){Villase\~{n}or}, {Vanderspek}, \&
  {the TESS Operations Team}}]{villasenor2019}
{Villase\~{n}or}, J., {Vanderspek}, R.~K., \& {the TESS Operations Team}. 2019,
  in {TESS Science Conference I}.
\newblock \url{https://tsc.mit.edu/2019/posters/Villasenor\_Joel.pdf}

\bibitem[{{Vokrouhlick{\'y}} {et~al.}(2015){Vokrouhlick{\'y}}, {Bottke},
  {Chesley}, {Scheeres}, \& {Statler}}]{2015aste.book..509V}
{Vokrouhlick{\'y}}, D., {Bottke}, W.~F., {Chesley}, S.~R., {Scheeres}, D.~J.,
  \& {Statler}, T.~S. 2015, in Asteroids IV, ed. P.~{Michel}, F.~E. {DeMeo}, \&
  W.~F. {Bottke}, 509--531, \dodoi{10.2458/azu_uapress_9780816532131-ch027}

\bibitem[{{Woods} {et~al.}(2021){Woods}, {Ruprecht}, {Kotson}, {Main}, {Evans},
  {Varey}, {Vaillancourt}, {Viggh}, {Brown}, \& {P{\'a}l}}]{woods2021}
{Woods}, D.~F., {Ruprecht}, J.~D., {Kotson}, M.~C., {et~al.} 2021, \pasp, 133,
  014503, \dodoi{10.1088/1538-3873/abc761}

\bibitem[{Zonca {et~al.}(2019)Zonca, Singer, Lenz, Reinecke, Rosset, Hivon, \&
  Gorski}]{Zonca2019}
Zonca, A., Singer, L., Lenz, D., {et~al.} 2019, Journal of Open Source
  Software, 4, 1298, \dodoi{10.21105/joss.01298}

\end{thebibliography}
\bibliographystyle{aasjournal}

\BeforeBeginEnvironment{appendices}{\clearpage}
\begin{appendix}

\section{\small{Possible pole orientation solutions for the four selected asteroids}} \label{App:figs}

\begin{figure}[!ht]
    \centering
    \rotatebox{-90}{
        \begin{minipage}{0.5\textwidth}
            \centering
            \begin{minipage}{\textwidth}
                \centering
                \includegraphics[scale=0.45]{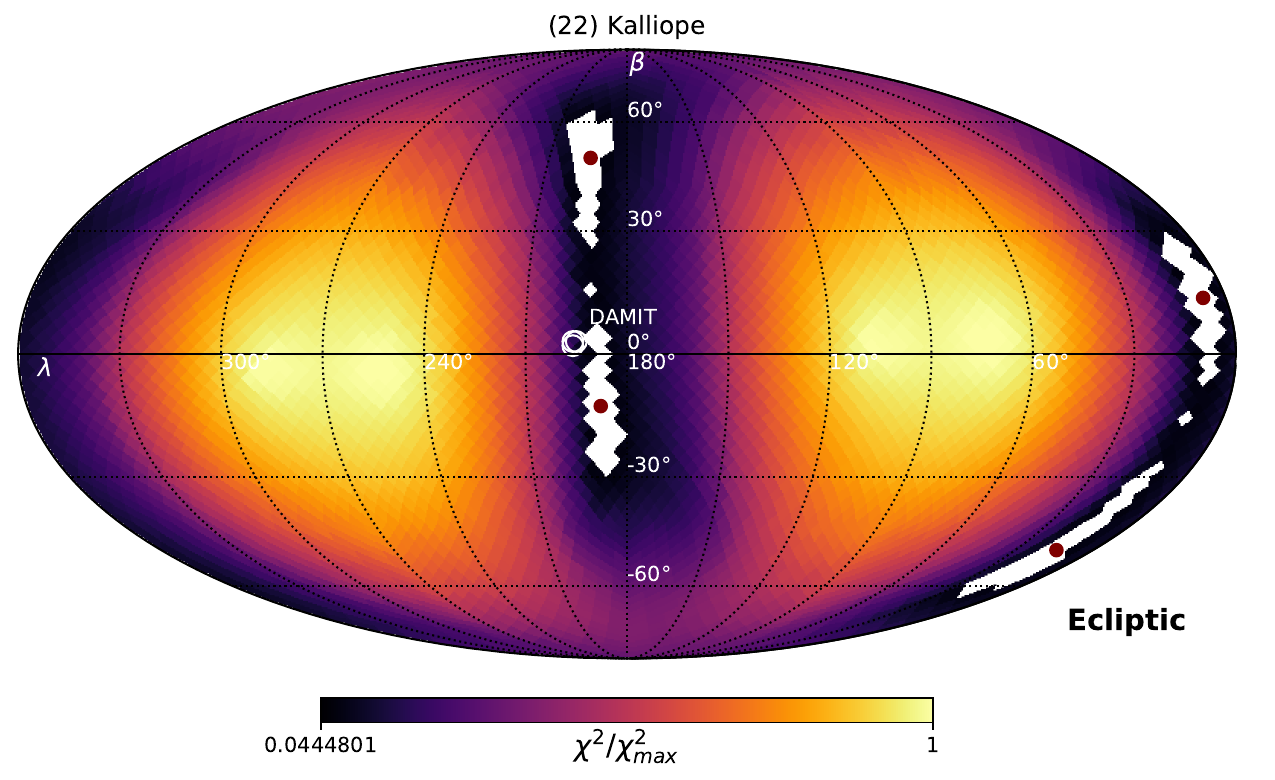}
                \label{fig:22_rot_dirs_lommelSeeliger_tess}
            \end{minipage}
            \begin{minipage}{\textwidth}
                \centering
                \includegraphics[scale=0.45]{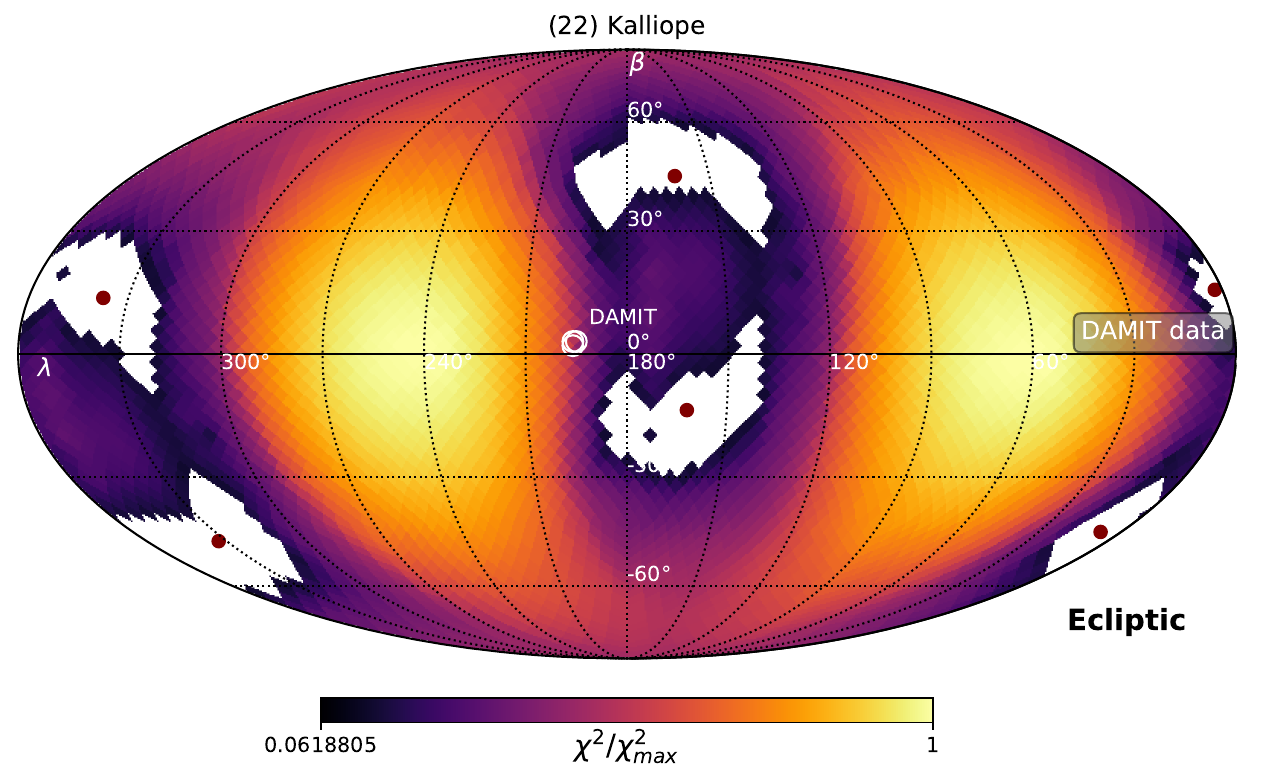}
                \label{fig:22_rot_dirs_lommelSeeliger_damit}
                \caption{$\chi^2$ contour maps of possible pole orientations for TESS light curves (top panel) and DAMIT light curves (bottom panel), for the asteroid (22) Kalliope. The white pixels show the smallest 2\% values of the data while the red dot shows their median. White circles represent the DAMIT pole solutions.}
            \end{minipage}
            \label{fig:22_combined_figure}
        \end{minipage}
    }
    \label{fig:22}
\end{figure}

\begin{figure}[htb]
    \centering
    \rotatebox{-90}{
        \begin{minipage}{0.5\textwidth}
            \centering
            \begin{minipage}{\textwidth}
                \centering
                \includegraphics[scale=0.45]{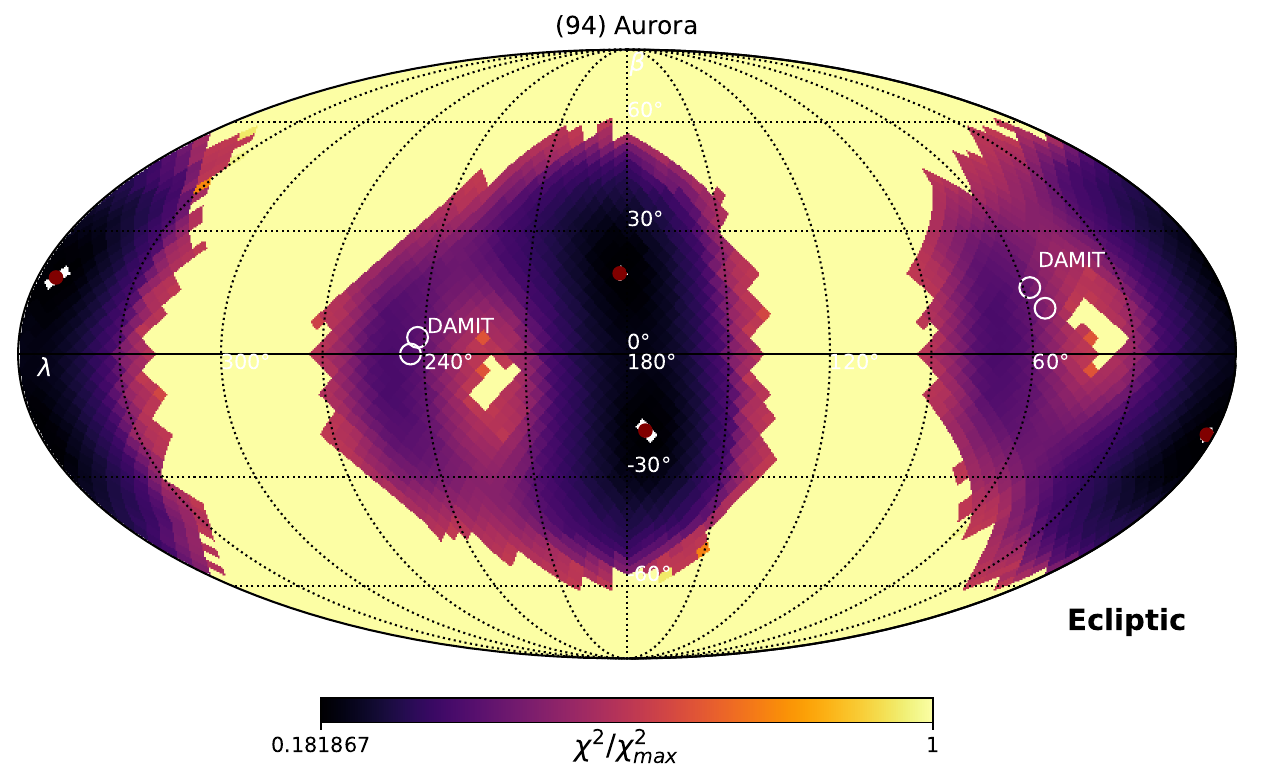}
                \label{fig:94_rot_dirs_lommelSeeliger_tess}
            \end{minipage}
            \begin{minipage}{\textwidth}
                \centering
                \includegraphics[scale=0.45]{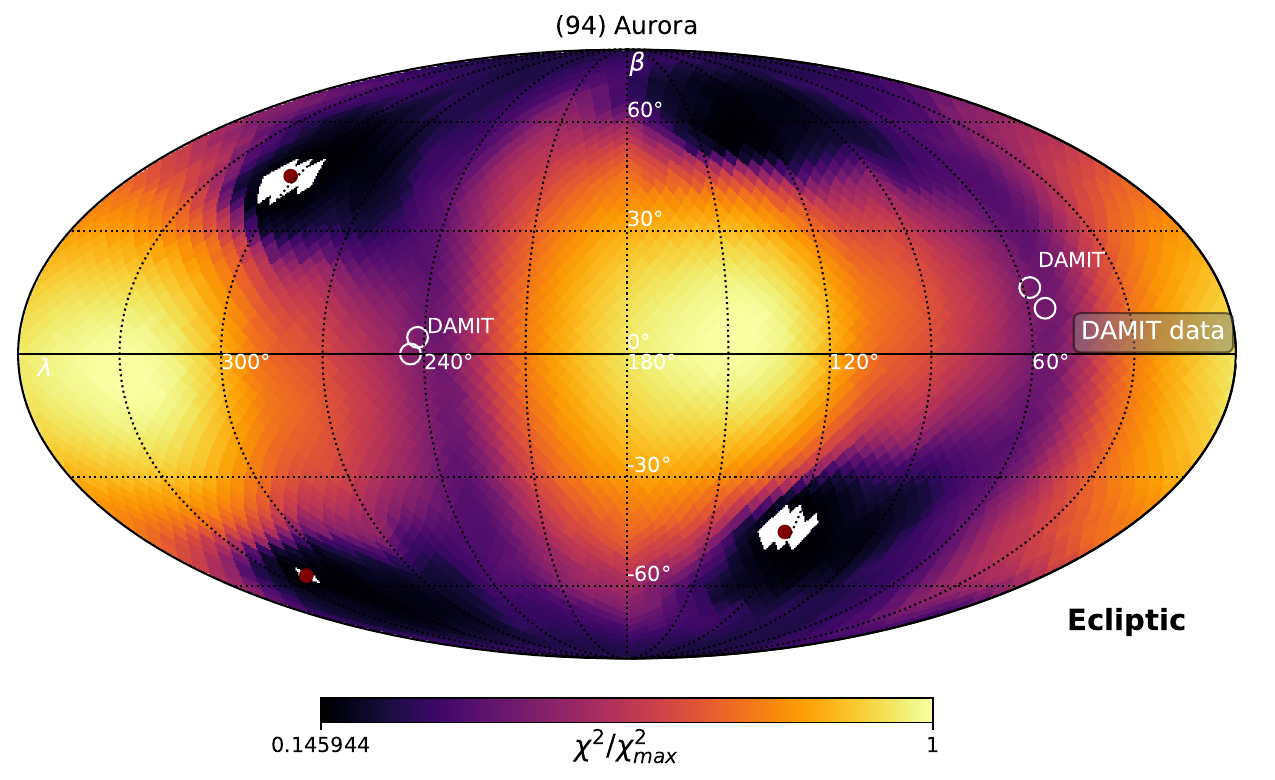}
                \label{fig:94_rot_dirs_lommelSeeliger_damit}
                \caption{$\chi^2$ contour maps of possible pole orientations for TESS light curves (top panel) and DAMIT light curves (bottom panel) for the asteroid (94) Aurora. The white pixels show the smallest 2\% values of the data while the red dot shows their median. White circles represent the DAMIT pole solutions.}
            \end{minipage}
            \label{fig:94_combined_figure}
        \end{minipage}
    }
\end{figure}

\begin{figure}[htb]
    \centering
    \rotatebox{-90}{
        \begin{minipage}{0.5\textwidth}
            \centering
            \begin{minipage}{\textwidth}
                \centering
                \includegraphics[scale=0.45]{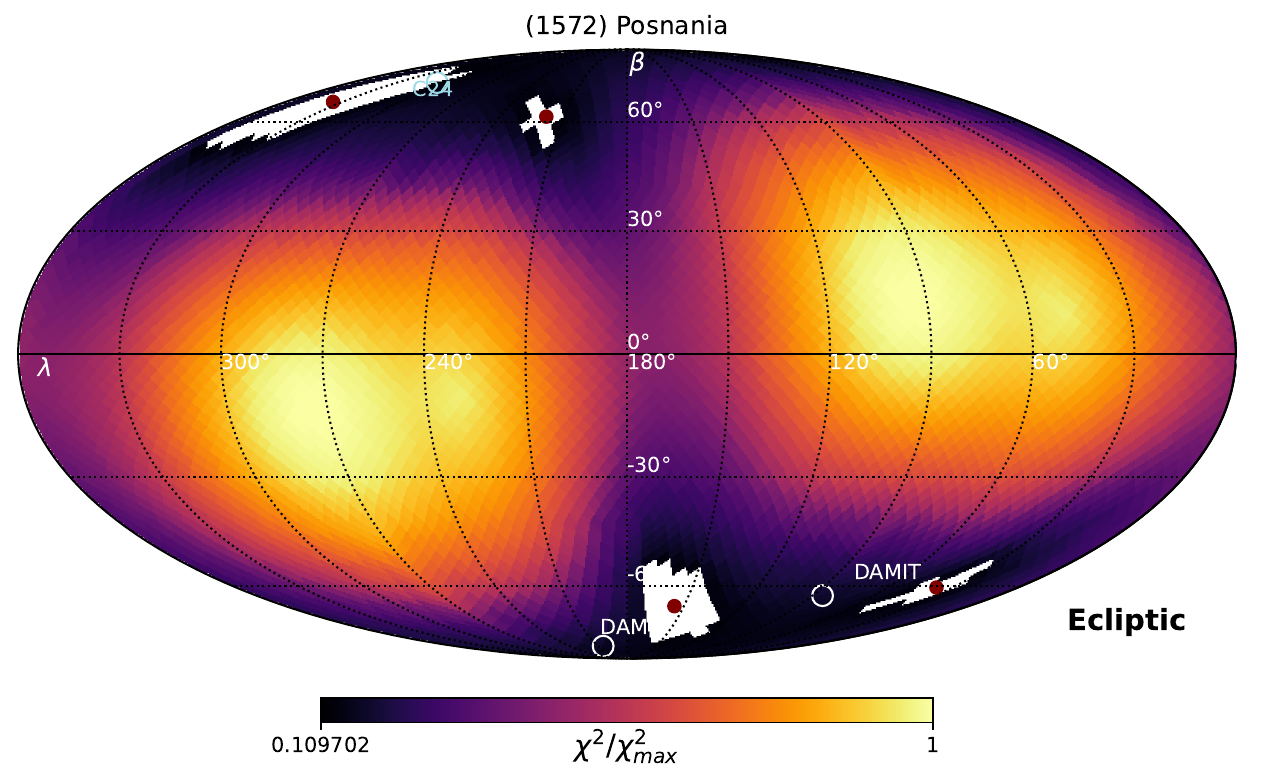}
                \label{fig:1572_rot_dirs_lommelSeeliger_tess}
            \end{minipage}
            \begin{minipage}{\textwidth}
                \centering
                \includegraphics[scale=0.45]{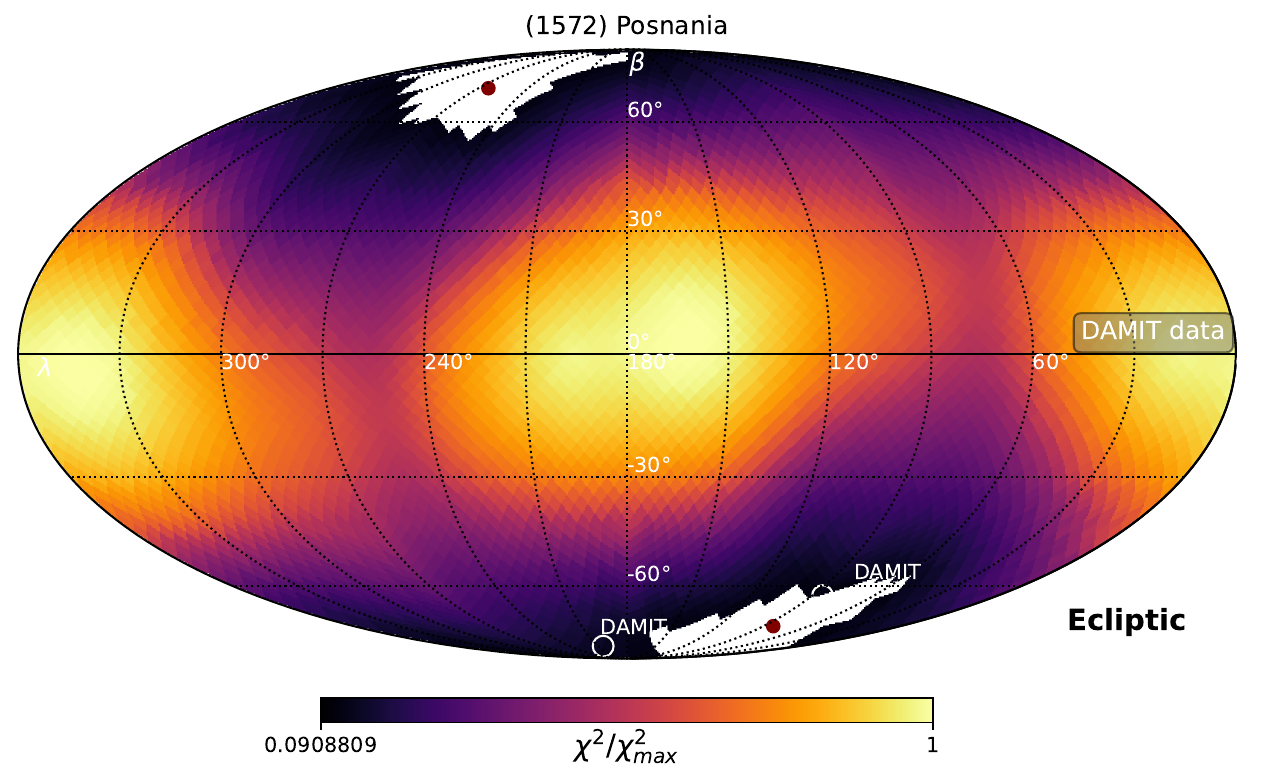}
                \label{fig:1572_rot_dirs_lommelSeeliger_damit}
                \caption{$\chi^2$ contour maps of possible pole orientation for TESS light curves (top panel) and DAMIT light curves (bottom panel) for the asteroid (1572) Posnania. The white pixels show the smallest 2\% values of the data while the red dot shows their median. White circles represent the DAMIT coordinates. The light blue circle on the top panel represents the C24 solution.}
            \end{minipage}
            \label{fig:1572_combined_figure}
        \end{minipage}
    }
\end{figure}

\end{appendix}

\end{document}